\documentclass[10pt]{article}

\usepackage[utf8]{inputenc}
\usepackage{url}
\usepackage{fullpage}
\usepackage{amsmath,amsfonts}
\usepackage{graphicx}
\usepackage{subcaption}
\usepackage{natbib}
\usepackage{color}
\title{The Minimum Wage as an Anchor: \\ {Effects on Determinations of Fairness by Humans and AI}\thanks{Submitted to the 2022 Regeneron Competition while the author was at Polytechnic School, 1030 E. California Blvd., Pasadena, CA 91106.}} 
\author{Dario G. Soatto \\~\\
Stanford University \\ %1450 Jane Stanford Way, Stanford, CA, 94305, USA \\ 
{\tt \small soatto@stanford.edu}\\ 
{\scriptsize {\tt https://www.linkedin.com/in/dario-soatto-70911b20a/}}
}
\date{August 28, 2022 \\~\\ 
Current Revision: November 11, 2022
\\~\\ {\small {\bf Keywords:} Neuroeconomics, Minimum Wage, Anchoring Effect, Crowdsourcing, Artificial Intelligence, AI Fairness, AI Bias, AI Equity, AI Explainability, GPT-3, Paraphrasing, \\ Artificial Behavioral Economics, Artificial Neuroeconomics.} 
}

\begin{document}

\titlepage

\maketitle

%\thispagestyle{empty}

%\newpage

\begin{abstract}
I study the role of minimum wage as an anchor for judgements of the fairness of wages by both human subjects and artificial intelligence (AI), specifically large language models (LLMs). Through surveys of human subjects enrolled in the crowdsourcing platform Prolific.co and queries submitted to OpenAI's LLM (GPT-3), I test whether the numerical response for what wage is deemed fair for a particular job description changes when respondents and GPT-3 are prompted with additional information that includes a numerical minimum wage, whether realistic or unrealistic, relative to a control where no minimum wage is stated. I find that the minimum wage influences the distribution of responses for the wage considered fair by shifting the mean response toward the minimum wage, thus establishing the minimum wage's role as an anchor for judgements of fairness. However, for unrealistically high minimum wages, namely \$50 and \$100, the distribution of responses splits into two distinct modes, one that approximately follows the anchor and one that remains close to the control, albeit with an overall upward shift towards the anchor. The anchor exerts a similar effect on the LLM; however, the wage that the LLM perceives as fair exhibits a systematic downward shift compared to human subjects' responses. For unrealistic values of the anchor, the responses of the LLM also split into two modes but with a smaller proportion of the responses adhering to the anchor compared to human subjects. As with human subjects, the remaining responses are close to the control group for the LLM but also exhibit a systematic shift towards the anchor. During experimentation, I noted some variability in the LLM responses depending on small perturbations of the prompt, so I also test variability in the bot's responses with respect to more meaningful differences in gender and race cues in the prompt, finding anomalies in the distribution of responses.

\end{abstract}

%\thispagestyle{empty}

%\newpage

%\newpage

%\setcounter{page}{1}

\begin{figure}[h]
\linespread{1}
\begin{center}
\includegraphics[width=.8\textwidth,height=.5\textwidth]{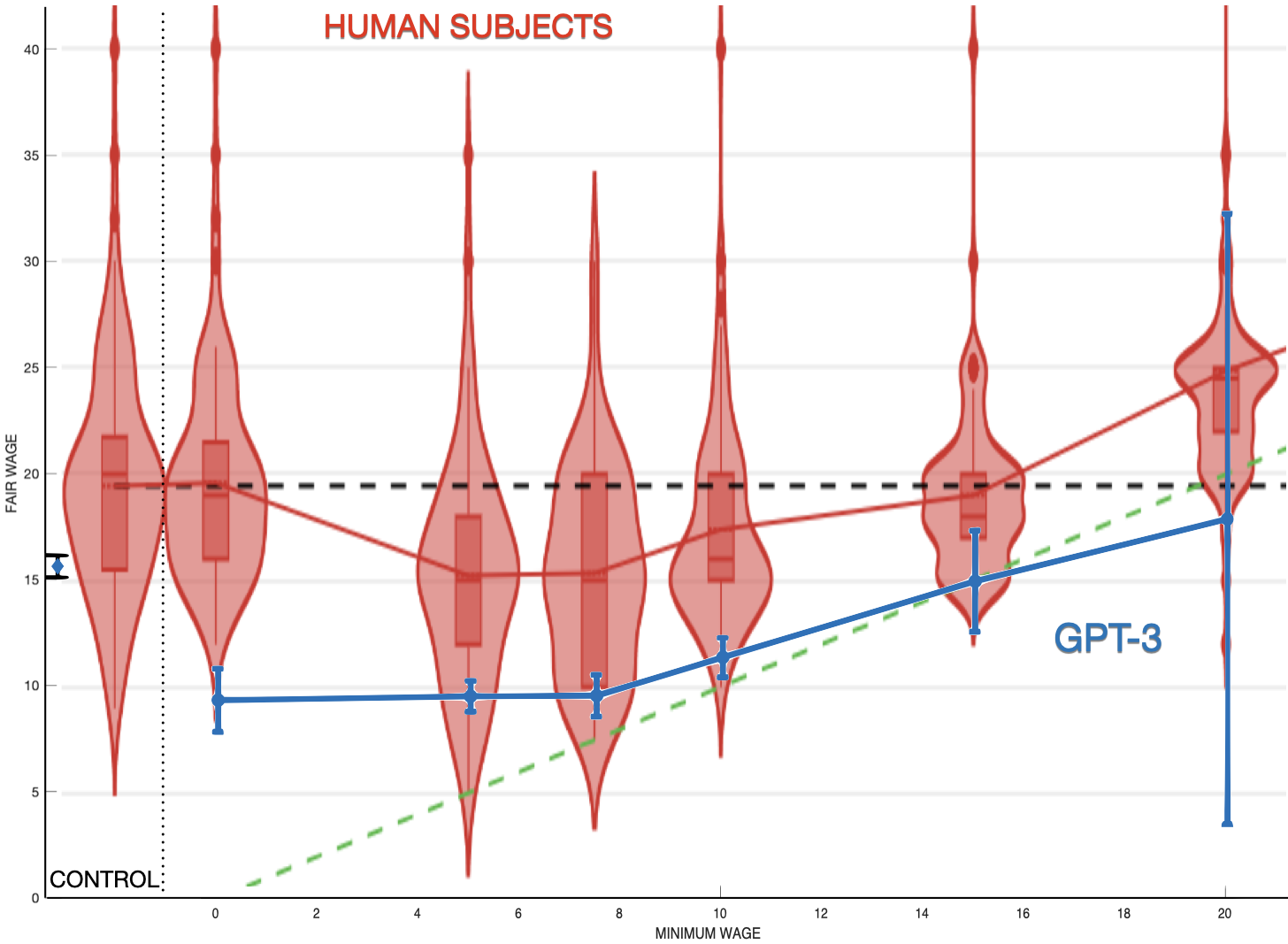}
\end{center}
\vspace{-.3cm}
\caption{{\bf Minimum Wage acts as an anchor on humans' and AI's determination of what constitutes a fair wage.} The distribution of what human respondents consider a fair wage for a materials worker depending on the stated minimum wage is depicted by the violin plot in red. The mean and three standard deviations of the responses from GPT-3 are shown in blue. Left of zero is the control, showing the distribution of responses when no minimum wage is stated in the prompt. The black dashed line shows the mean control response, and the green dashed line provides a linear reference for where the responses would appear if they exactly equal the minimum wage. Note the systematic shift between human workers and GPT-3, on average \$5.32, possibly due to the outdated nature of data used to train GPT-3. The same plots, with GPT-3's responses shifted up to align the responses, is shown in Fig.~\ref{fig:aligned} (right), along with the response with aligned controls (left).
}
\label{fig:splash}
\end{figure}

\section{Introduction}
Behavioral economics studies how decisions, particularly economic decisions, are affected by judgemental heuristics (\cite{belsky2010smart}). Judgemental heuristics are rules of thumb that help \textit{``reduce the complex tasks of assessing probabilities and predicting values to simpler judgmental operations"} (\cite{tversky1974judgment}). Although judgemental heuristics allow people to make complex and difficult decisions rapidly, they often result in consistent and predictable biases (\cite{tversky1974judgment}). Among these heuristics is the anchoring effect, which is both robust and pervasive in decision-making (\cite{furnham2011literature}).

The anchoring effect refers to {\em ``a systematic influence of initially presented numerical values on subsequent judgments of uncertain quantities," where the judgement is biased toward the anchor} (\cite{teovanovic2019individual}). The anchoring effect has been replicated across a variety of contexts, as I discuss in Sect.~\ref{sec:related}, including with judgements involving money and anchors established by government policy. Given the prevalence of the anchoring effect, one would expect to find an anchoring bias generated by one of the most controversial figures of the economy: the minimum wage.

Since determining what wage to offer to employees is a complex judgement, employers may use the minimum wage as a convenient reference point upon which to base their offers. Due to the difficulty of conducting controlled experiments with employers, I seek to answer a related question: does the minimum wage function as an anchor for what people perceive to be a fair wage? Although the average person does not engage in wage determination, public discourse surrounding the fairness of wages can influence the economy through the political process, meaning that effects of minimum wage on perceptions of fairness can have broad implications. Thus, I ask human subjects on the crowdsourcing platform Prolific.co as well as an AI bot, specifically a version of OpenAI's bot GPT-3 (\cite{davinci}), to determine the fair wage for a worker given their job description and a minimum wage. 

In addition, I explore paraphrasing techniques to test the stability and robustness of the AI bot's responses in the face of perturbations of the input, and I conduct side experiments to evaluate how the presence of cues for protected attributes of the worker, namely gender and race, affects the AI's responses. 

In the next section, I survey the related work before presenting my methods and findings in the subsequent two sections.

\subsection{Related work}
\label{sec:related}

Anchoring in decision-making was first proposed by Paul Slovic in researching how people evaluate the risk of gambling (\cite{slovic1967relative}). Subsequent research has expanded the literature on the anchoring effect, replicating it across a wide variety of contexts. In a pioneering study, Tversky and Kahneman spun a wheel of fortune in front of participants and asked them to consider whether the number of African countries in the United Nations was higher or lower than the random outcome (a prior anchoring question) before asking for their final estimation of the number. The authors found that the final estimation was significantly affected by the value of the initial anchor (\cite{tversky1974judgment}). Similar prior anchoring questions have been found to influence the certainty equivalent for a gamble (\cite{johnson1989bias}), estimation of tree height (\cite{jacowitz1995measures}), and willingness to subject oneself to annoying sounds for money (\cite{ariely2003coherent}). Anchoring effects, however, are also present for consequential decisions, both economic and non-economic. Judges' sentencing decisions and bankruptcy rulings are anchored by prosecutors' demands (\cite{enough2001sentencing}) and official reviewers' recommendations (\cite{mugerman2021courts}), respectively. Moreover, arbitrarily generated anchors have been found to affect willingness to pay for public goods (\cite{green1998referendum,kahneman1993stated}), willingness to pay for private goods (\cite{ariely2003coherent}), willingness to pay to save seabirds (\cite{jacowitz1995measures}), and judges' sentencing (\cite{englich2006playing}); indeed, the anchoring effect is present even when participants know that the anchor is generated by their own social security numbers (\cite{ariely2003coherent}) or dice rolls (\cite{englich2006playing}).

Anchoring effects have been observed broadly (\cite{mussweiler2001semantics}), including on judgments involving money: for example, assessments of real estate value are significantly affected by the asking prices, even for professional real estate agents who take pride in ignoring the asking price (\cite{northcraft1987experts}). Anchoring has also been studied in the context of government policy. For instance, studies have explored how damage caps on civil suits affect the ability of parties to reach a settlement (\cite{pogarsky2001damage}).

Anchoring effects have been found to be difficult to avoid: they occur even when people are instructed to ignore the anchor (\cite{wilson1996new}), and they can influence decision-making even with a one week delay between considering the anchor and making the final judgement (\cite{mussweiler2005subliminal}).

There are two main theoretical explanations of anchoring: the anchoring and adjustment heuristic and suggestion. The anchoring and adjustment heuristic was proposed in Tversky and Kahneman in their study of judgement under uncertainty (\cite{tversky1974judgment}). They theorized that the anchoring effect stems from insufficient adjustment away from the anchor: people consider the anchor as a starting point and adjust away from it until they reach a ``zone of uncertainty," resulting in insufficient adjustment because adjusting away from the anchor is a conscious and effortful process. In support of this theory, it has been found that anchoring is more pronounced when mental resources are depleted, whether due to loaded memory or inebriation, and nodding in response to an anchor results in a greater anchoring effect (\cite{epley2001putting}).

An alternative theory of anchoring advanced by Strack and Mussweiler postulates that anchors create a priming effect: people test the possibility that the anchor value reflects the true value and construct a mental model that includes information consistent with the anchor value (\cite{strack1997explaining}). Accordingly, when people are asked to estimate the average price of German cars, high anchors selectively prime the names of luxury brands while low anchors prime brands associated with mass-market cars (\cite{mussweiler1999hypothesis}), and asking people to imagine the opposite of the anchor mitigates the anchoring effect (\cite{mussweiler2000overcoming}).

The possibility that public policy generates anchoring effects has not been overlooked by the behavioral science literature. It has been found that imposing mandatory minimum annuity laws for retirement savings (\cite{hurwitz2020unintended}), limiting payment-to-income ratios and maturities in mortgage markets (\cite{mugerman2020anchoring}), and minimum sick pay provisions (\cite{Bauernschuster2010mandatory}) generate anchoring effects.

However, based on my search of the literature, I believe this is the first study to measure the effect of minimum wage on perceived fairness of wages associated with a given job description. I am also not aware of any previous measurements of anchoring effects in AI Bots. The results of my experiments, which show the population of responses splits into multiple nodes in the presence of unrealistic anchors, also point to limitations in the traditional analysis of anchoring effects.

The research in the literature most similar to mine that I have found is a 2012 study on how symmetrical and asymmetrical knowledge of minimum wage among workers and employers can influence the wages offered (\cite{wang2012workers}); however, there is no mention of the anchoring effect as the genesis of this behavioral phenomenon. Meanwhile, in a comment titled ``The Cognitive Effect of a Minimum Wage" Engel speculates that the anchoring effect may have played a role in laboratory experiments of a simulated labor market with a minimum wage performed in (\cite{engel2007cognitive, falk2006epl}). My study aims to substantiate this hypothesis.

In contrast to most anchoring research, this study examines the anchoring effect through online crowdsourcing methods. The validity of such polling data has been verified (\cite{goodman2013data, meyers2020reducing}), and classical anchoring experiments have been reproduced using Amazon Mechanical Turk, a crowdsourcing platform (\cite{wolfson2013assessment}). 

As described in (\cite{wolfson2013assessment}, Sect.~III.C), the authors conducted the classical experiment measuring the willingness-to-pay for a bottle of wine, which they found difficult to replicate due to reasons such as {\em ``the experiment relies upon subjects seeing an actual bottle of wine, rather than an image''} and {\em ``the fabricated wine description may have been transparently fabricated and/or a poor description of wine''}. The authors were able to replicate the anchoring experiments of (\cite{stewart2009cost} ) on credit card minimum payments, showing that {\em ``subjects shown the minimum payment were just as likely to pay off the entire balance, but otherwise made significantly lower payments. Those not shown the minimum paid on average almost double those that were shown the minimum. The evidence suggests that people arbitrarily anchor on the minimum payment if given the choice.''}

While the anchoring effect had already been replicated on crowdsourcing platforms, I tested one of the original anchoring experiments on the bot GPT-3. In particular, the question  {\tt "What is the percentage of African Countries in the United Nations?"} after a prompt {\tt "Is the percentage of African countries in the United Nations higher or lower than 10\%? Answer higher or lower}" and obtained a top answer of {\tt 9\%} with probability 77.27\%,  {\tt 8\%} with probability 4.92\%, {\tt 7\%} with probability 2.17\% and {\tt 5\%} with probability 1.38\%. However, changing the prompt to a high anchor ({\tt Is the percentage of African countries in the United Nations higher or lower than 65\%? Answer higher or lower}) yielded answers to the same question of {\tt 54\%} with probability 25.28\%, {\tt 63\%} with probability 9.67\% and {\tt 64\%} with probability 3.74\%. 

When using the AI bot GPT-3 (\cite{brown2020language}), I generate a distribution of responses by exploiting the probabilities produced by the model, as explained in Sect.~\ref{sec:controlling}. A single response from GPT-3 aggregates information from multiple sources and produces output words by sampling words according to the softmax probability produced by the Transformer (\cite{vaswani2017attention}). This allows me to study the variability in response to the same prompt with a single run of the experiment. After I completed the experiment and described it in an earlier draft, I was made aware of other recent work using an AI bot in lieu of surveys (\cite{argyle2022out,aher2022using}) although not in the context of anchoring effects. 

Variability in response to different prompts has allowed scientists to use GPT-3 to evaluate the biases of AI Bots (\cite{aher2022using}) including, recently, biases in job descriptions (\cite{borchers2022looking}). Others have looked at the effect of gendered speech (\cite{lucy2021gender}) and more general issues in cognitive psychology (\cite{binz2022using}). In this project, I do not engage in analysis specific to GPT-3; however, I conduct pointed experiments to assess the stability of GPT-3's responses with respect to grammatical, syntactic and semantically meaningful perturbations.

If AI studies the development of intelligent traits in a machine being trained, studying the behavioral decision-making of the trained machine can be considered an exercise in ``Artificial Behavioral Economics.’’ or ``Artificial Neuroeconomics.'' While such disciplines do not exist yet, I expect research in these areas to become increasingly important as machine learning systems become part of economic decision-making. 

\subsection{Summary of Outcomes}

I test human subjects through the crowdsourcing service {\tt Prolific.co} as well as AI Bots such as {\tt BlenderBot} 
and {\tt GPT-3}. I refer to the subjects polled through Prolific.co as ``Prolific workers."

\begin{figure}[h]
\linespread{1}
\begin{center}
\includegraphics[width=.49\textwidth,height=.3\textwidth]{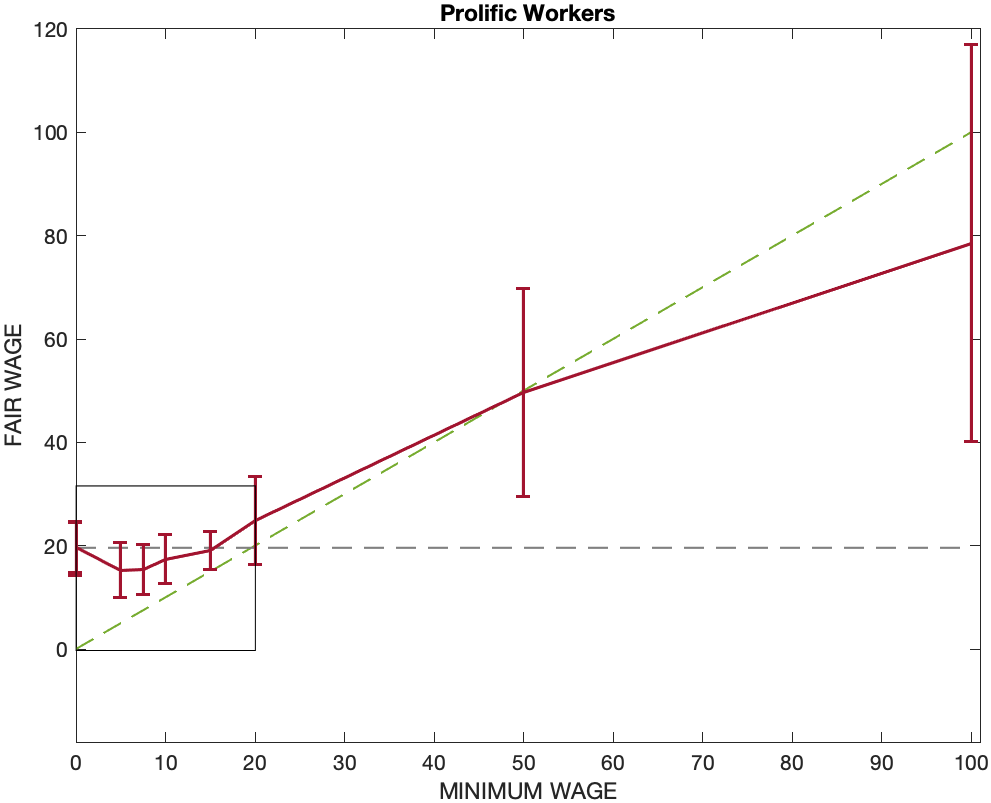}
\includegraphics[width=.49\textwidth,height=.3\textwidth]{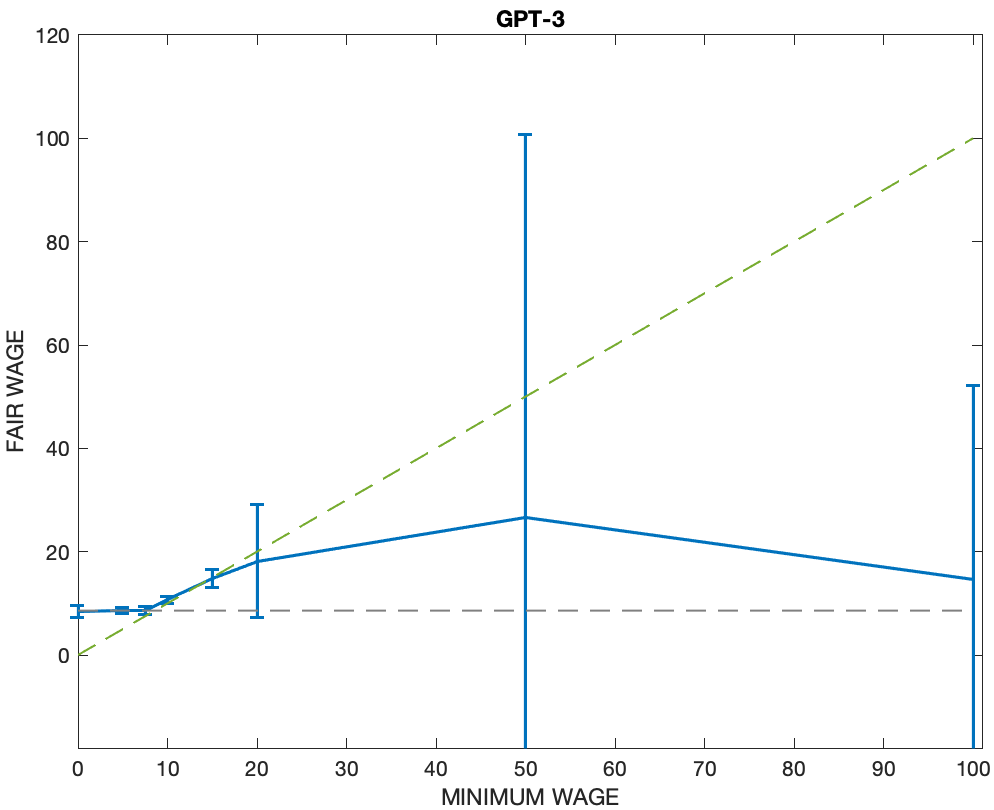}\\
\includegraphics[width=.49\textwidth,height=.3\textwidth]{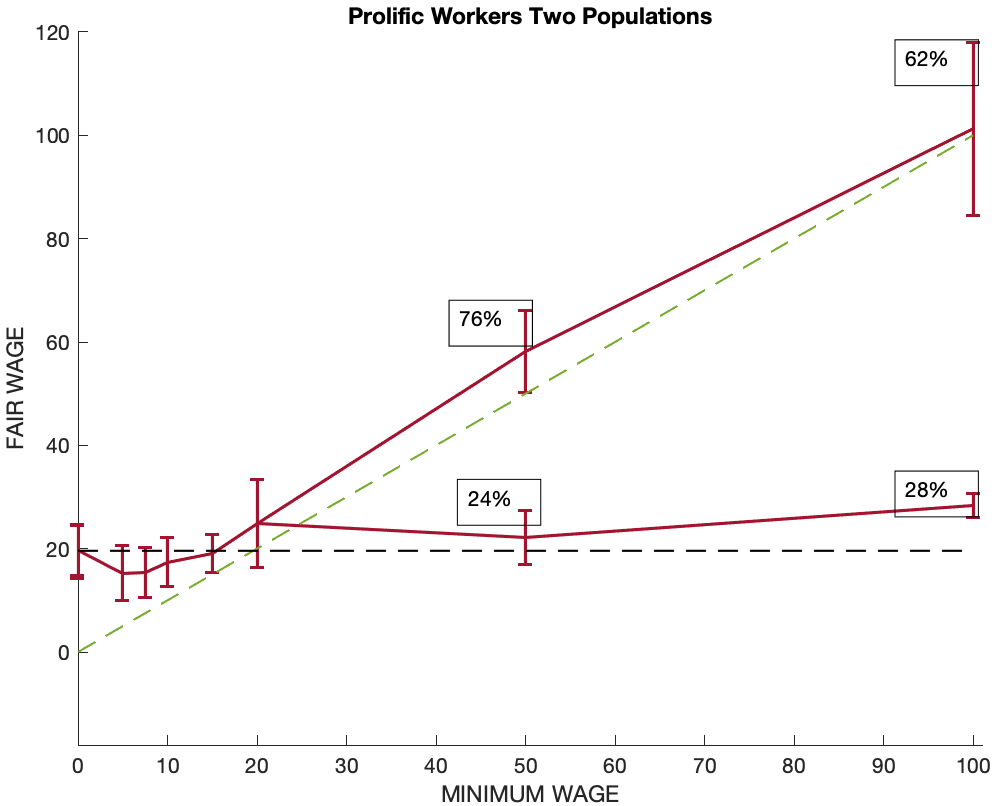}
\includegraphics[width=.49\textwidth,height=.3\textwidth]{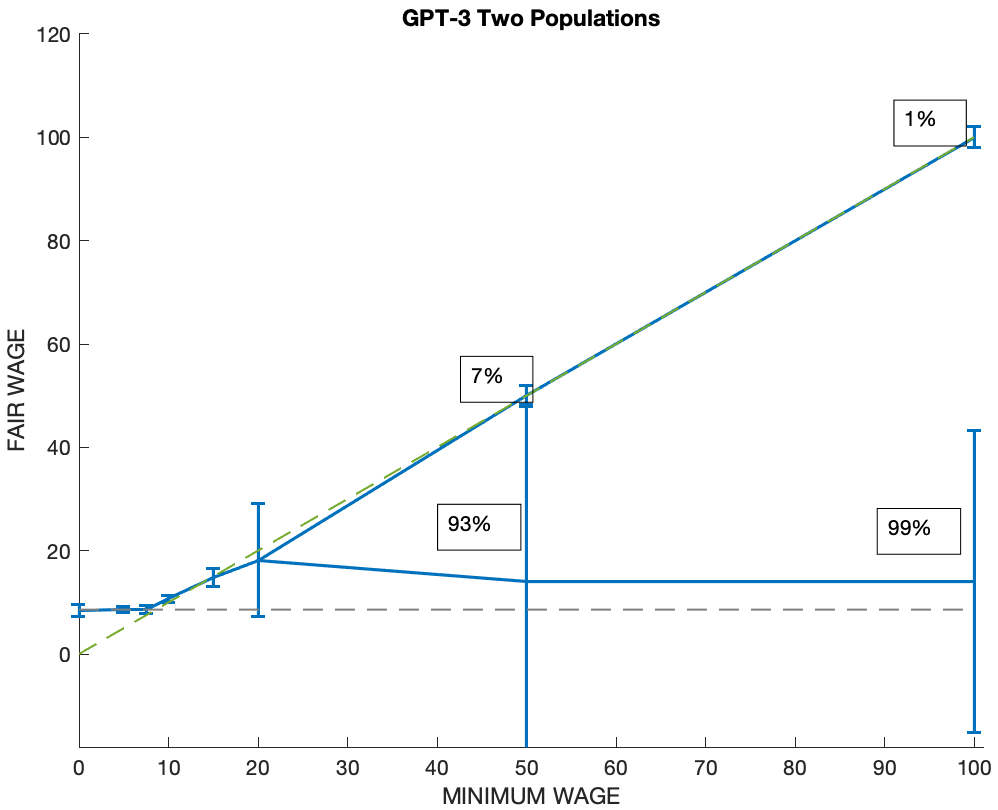}
\end{center}
\caption{{\bf Populations Split for Unrealistic Anchors. } For unrealistic anchors, the responses of Prolific workers (left) split into two populations (bottom left), with the majority of respondents following the anchor and the rest remaining closer to the control but with a systematic upward shift. For GPT-3 (right), however, the two populations are distributed differently from the Prolific Workers, with a small percentage of responses following the unrealistic anchor and the majority remaining close to the control (bottom-right). Fig.~\ref{fig:splash} corresponds to the inset in the top-left plot. The left column can be visualized as a violin plot based on population statistics, as shown in Fig.~\ref{fig:violin}.
}
\label{fig:bimodal}
\end{figure}

\begin{itemize}

\item When there is no mention of the minimum wage, the wage perceived as fair is significantly higher than the current minimum wage in any American city or state. Stating that there is no minimum wage does not significantly change what is perceived as a fair wage. 
\item I demonstrate that the minimum wage functions as an anchor for what Prolific workers consider a fair wage: for numerical values of the minimum wage ranging from \$5 to \$15, the perceived fair wage shifts towards the minimum wage, thus establishing its role as an anchor (Fig.~\ref{fig:splash} and Table~\ref{tab:summary}). I replicate this result for a second job description, finding that the effect holds even for jobs where wages are supplemented by tips.
\item For unrealistic values of the stated minimum wage, namely \$50 and \$100, I find that first-order analysis is insufficient as the distribution of responses splits into two distinct modes: one closely follows the anchor while the other remains closer to the control, but still with a systematic bias towards the anchor (Fig.~\ref{fig:bimodal}). I characterize the value of the anchor where the two populations split using the notion of the resolving power of a lens. 
\item For the AI Bot, I find that it also exhibits anchoring effects but differs from Prolific workers in important ways. First, for realistic anchors, the AI bot exhibits an overall downward offset in the fair wage response. I hypothesize that this may be due to the time when the data used for training GPT-3 was collected. Second, for unrealistic anchors, GPT-3 predominantly favors the mode of the distribution closer to its responses when given realistic wages, rather than blindly following the anchor value.
\item In order to obtain a diversity of responses from GPT-3, I explore different sources of variability including sampling from the output probabilities and using paraphrased prompts. I observe that the distribution of responses changes as a result of modest changes of the inputs including punctuation, spelling, and insertion of cues to protected attributes such as gender and race (Table~\ref{fig:parrot}).
\item I find mild variation among responses for different perturbations but strong effects associated with changing gendered pronouns (Table~\ref{fig:bias}): changing the pronoun from ``he'' to ``she'' caused the perceived fair wage to change from \$15 to \$12. However, identical tests conducted at a later time showed a decrease in the response for ``he'' from \$15 to \$12, matching the results with the pronoun ``she.'' I hypothesize that post-processing checks are put in place to equalize outcomes based on cues associated with protected attributes. I also test for different prompts individualized with race-stereotypical names and find small inconsistencies (Table~\ref{fig:race}). 
\item My analysis is limited to a full set of experiments for the most common job in the US, and spot-checks for a second. It is possible that the anchoring effect of the minimum wage may be different for higher-paying jobs. Perception of fairness of wages can also be influenced by additional factors such as age, geography, level of education, political inclination, health, and others, which I hope to explore in future studies.
\end{itemize}

\begin{figure}[h]
\linespread{1}
\begin{center}
\includegraphics[width=.9\textwidth]{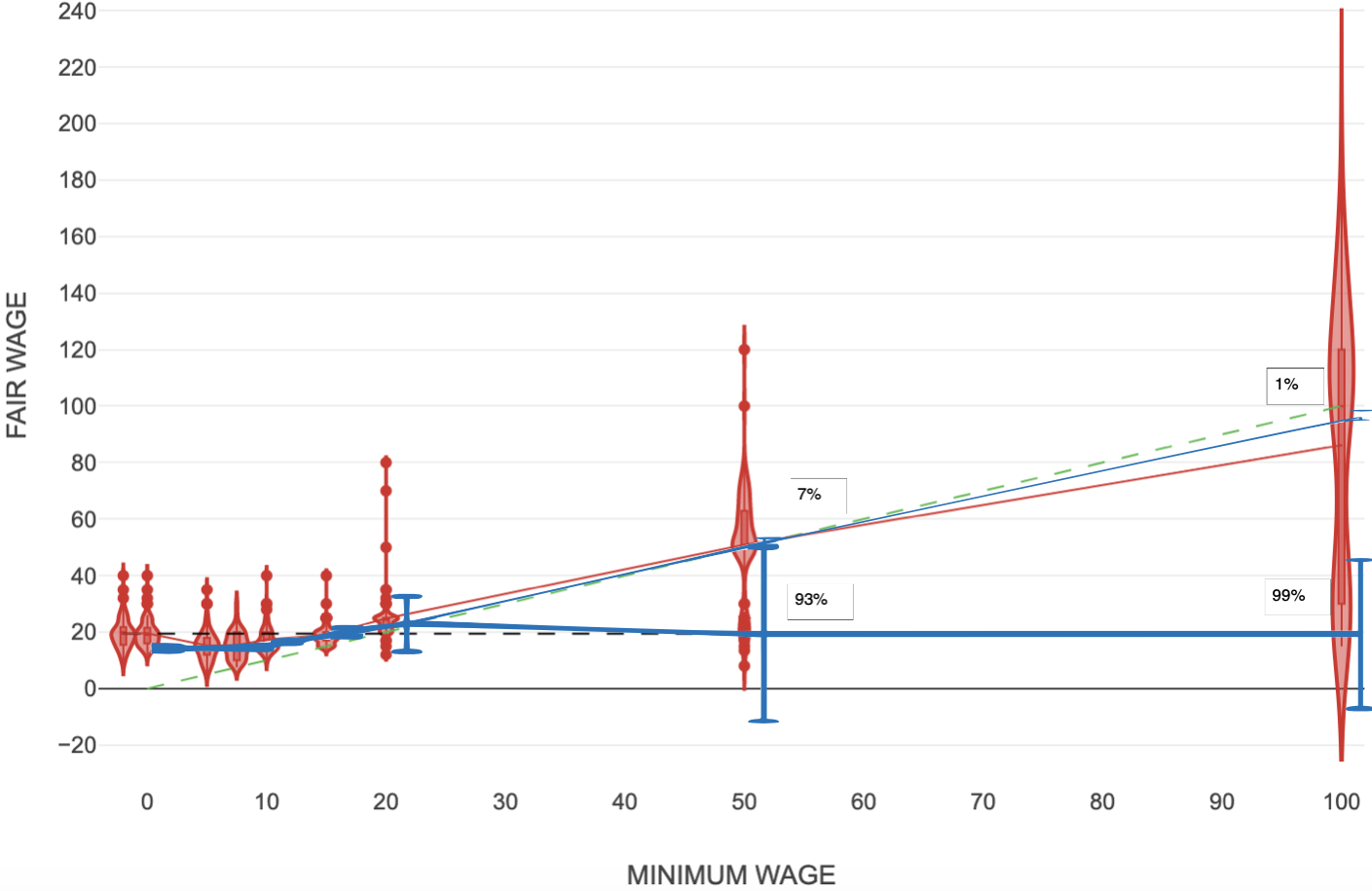}
\end{center}
\vspace{-.51cm}
\caption{{\bf Populations Split for Unrealistic anchors.}  The violin plot extends Fig.~\ref{fig:splash} for unrealistic anchors up to \$100.  The two distinct modes are seen corresponding to minimum wage \$50 and  \$100. The errorbar plot for GPT-3 is shifted to the right for visibility.
}
\label{fig:violin}
\end{figure}

\section{Method}

For the human component of my experiment, I explored using  Amazon GroundTruth and Prolific ({\tt prolific.co}) to recruit participants. For the automated component of my experiment, I explored Meta AI's BlenderBot and GPT-3, a large language model (LLM) developed by OpenAI that is the largest LLM publicly available through a graphical user interface. In the next two sections, I describe the specifics of each experimental setting.

\subsection{Human Subject Selection}

After investigating GroundTruth and Prolific.co, I opted to use Prolific.co to survey human subjects due to its high quality of responses and ease of use. Prolific is a platform for on-demand data collection focused on survey studies and experiments involving  paid human subjects. It has over 130,000 vetted participants who can be selected based on demographics, geography, language, education, and other attributes.

Prolific automates the selection of subjects based on an initial survey. I limited participation to residents of the United States without restriction to demographic characteristics, and I collected information about age, sex, ethnicity, country of birth, country of residence, nationality, language, student status, and employment status for possible future analysis. In accordance with Prolific.co's recommendations, I paid human subjects \$0.20 per response, which yields a wage of \$10-12/hour depending on individual response speed. This turned out to be lower than what GPT-3 deemed fair (Table~\ref{fig:race}). A sample enrollment form is included in the IRB documentation. The total cost of the experiment was \$425, which I funded from my personal savings. Polytechnic School IRB Approval of the protocol was obtained on 9/20/2022. 

\subsection{AI Bot Selection}

After experimenting with Meta's Blenderbot (\cite{blenderbot}) 
I settled on GPT-3  (\cite{davinci}).  Of the available models, I chose the largest one, which is the 175 billion-parameter model (DaVinci), with  default settings in the interface. Since the outcome of the model is randomized, as I discuss in Sect.~\ref{sec:controlling}, I submit the same prompt and query multiple times. In addition, for each query corresponding to job description and for each stated minimum wage, I generate additional queries via automated paraphrasing. This is done both to evaluate the stability of the answers provided by GPT-3 and to increase the diversity of responses to match the human population. I employ automated paraphrasing models available in open source through HuggingFace (\url{huggingface.co}). Specifically, I use a paraphrasing model called Parrot, which is based on T5, an model pre-trained by Google, (\cite{raffel2020exploring}) through a Python wrapper based on sample code from Github (\cite{parrot}). 

AI bot-generated answers are processed through the same protocol as human responses for the purpose of analysis. Both GPT-3 and Parrot are available for research purpose, and the experiments were conducted within the free tier of usage.

\subsection{Experimental Protocol}

Both human and AI respondents are first presented with one of two job descriptions selected for being the two most common jobs in the US according to the Bureau of Labor Statistics (\cite{occupational}): Materials Worker and Food and Beverage Serving Worker. I complete a full set of surveys using the job description for Materials Worker and a subset with Food and Beverage Serving Worker. All plots refer to data from surveys with the Materials Worker job description.
Once presented with a job description, a random subset of participants are given additional information about minimum wage $M$, which can take values $m = \{0, 5, 7.5, 10, 15, 20, 50, 100\}$ in US dollars. Information about the absence of minimum wage is represented by $m = 0$; this differs from the absence of information about the minimum wage.
Finally, each respondent is asked to indicate what they consider a fair wage for that job description. Their answer $R$ is recorded, with values ranging between $r = 0$ and $r = 50$. Improper answers were excluded from the survey and constituted less than 0.5\% of the responses.

A summary of the results for realistic values of the anchor is shown in Fig.~\ref{fig:splash}. The full range is in Fig.~\ref{fig:bimodal}. I aggregate data into histograms approximating the probabilities of a certain wage $P$ for each job description. For each job description, I compute:
\begin{equation}
P(R = r | M = m) 
\label{eq:conditional}
\end{equation}
for all response values $r$ and minimum wage values $m$, including $m=0$ (no minimum wage). The resulting histograms are compared with the control, which is the unconditioned distribution:
\begin{equation}
P(R = r) 
\label{eq:marginal}
\end{equation}
approximated by the frequency of responses $r$ obtained without mention of the minimum wage.

\subsection{Main Hypothesis}
\label{sec:formalization}
The main hypothesis is that minimum wage acts as an anchor on what is considered a fair wage for a job. That is, for a job description, the average response for what is considered a fair wage changes depending on whether it is conditioned on a value of the minimum wage $m$, and in particular it shifts towards that value $m$. As the minimum wage is decreased, I expect the wage that is deemed fair to decrease correspondingly. Next, I formalize this question into a statistical test.

If I call $\bar r_0$ the average response with respect to the distribution $P(R)$ in \eqref{eq:marginal} and $\bar r_m$ the average with respect to the conditional distribution $P(R|M)$ in \eqref{eq:conditional}, I have 
\begin{eqnarray}
\bar r_0  &=& \sum_{r=0}^{40} r P(R = r)  \nonumber \\ 
\bar r_m  &=&  \sum_{r=0}^{40} r P(R = r | M = m)
\end{eqnarray}
 and their standard deviations $\sigma_0  = \sqrt{\sum_r (r-\bar r_0)^2 P_j(R = r)}$ and  $\sigma_m = \sqrt{\sum_r (r-\bar r_m)^2 P_j(R = r| M = m)}$. I can now formalize the definition of anchoring into a statistical test: starting from the definition of anchoring in the introduction—``systematic \underline{influence} of initially presented {\em numerical values} ($M = m$) on subsequent judgments ($R = r_i)$ of uncertain quantities," where the judgement is \underline{biased toward} the anchor ($m$)—there are multiple ways to define ``influence'' and ``bias.'' Moreover, there are different ways to aggregate responses across individuals, leading to different ways of measuring the anchoring effect, which I describe next.

\subsection{Anchoring Scores}
\label{sec:scores}

The definition of anchoring leaves several options open to arrive at a quantitative measure of the effect, depending on how {\em influence} is quantified, how the {\em bias} toward the anchor is measured, and how the response of different individuals is aggregated.

The simplest case consists of {\em aggregating first} the responses of each individual with and without the anchor, $\bar r_m$ and $\bar r_0$ respectively, and then measuring the {\em influence} of the anchor by comparing the distance of the anchor from the control, $|\bar r_0 - m |$ and the distance of the anchor from the conditioned response  $|\bar r_m - m | $. If there is an anchoring effect, in general I expect the {\em Mean Anchoring} score to be positive: 
\begin{equation}
\bar \alpha_m = |\bar r_0 - m| - |\bar r_m - m| \ge 0
\label{eq:AMA}
\end{equation}
I can then obtain a numerical score by averaging over values of the anchor $m \in M_c$ where $M_c$ is the set in which I vary the anchor, specifically $M_c = \{0, 5, 7.5, 10, 15, 20\}$, obtaining 
what I call the {\bf Average Mean Anchoring} (AMA) score
\begin{equation}
\bar \alpha = \frac{1}{|M_c|} \sum_{m \in M_c} \bar \alpha_m 
\end{equation}
where $|M_c|$ denotes the number of elements in $M_c$. 
If there is an anchoring effect, $\bar \alpha$ has to be greater than zero. 

Mean analysis, whether with AMA or another score, is only sound if the distribution of responses is unimodal (i.e. the mean is representative of the data. If the responses split into multiple modes, then linear analysis is no longer appropriate. Since the response data do in fact show a splitting of the distribution, (Fig.~\ref{fig:bimodal}) in order to capture the potential multi-modality of responses, I also devise a {\em distributional distance} $D$ between the distribution that yields $\bar r_m$, which is $P(R|M = m)$, and the one that yields $\bar r_0$, which is $P(R)$. If I call the distance between two distributions $D$, I can define
\begin{equation}
\delta_m = D\left(P(R|M=m),  P(R)\right)
\end{equation}
and what can be termed the {\bf Average Distributional Anchoring} score (ADA)
\begin{equation}
\delta = \frac{1}{|M_c|} \sum_{m \in  M_c} \delta_m.
\end{equation}
In particular, choosing $D(P_0,P_m) = \sum P_m \log{P_m/P_0}$, with $P_0 = P(R = r)$ and $P_m = P(R = r | M = m)$ leads to $\delta$ measuring the {\em Shannon Mutual Information} $I(R, M)$ between the random variables $R$ and $M$: 
\begin{equation}
\delta = K %~ |M_c|~ 
I(R, M) %=  \frac{1}{|M_c|}\sum_{m \in M_c}\frac{1}{K} \sum_{r = 0}^{K} P(R=r|M=m) \log\frac{P(R=r | M=m)}{P(R=r)} % since P(m) = const = 1/|M_c|
\label{eq:ADA}
\end{equation}
with $K$ being the number of bins in the histogram (\cite{cover1999elements}). To avoid division by zero in the logarithm, I add a unit of mass to each bin and re-normalize. Mutual Information is measured in  BITs or NATs depending on the choice of logarithm base (binary or natural). For the analysis in this paper, I use the natural base.  

The advantage of Mutual Information is that its value can be interpreted as the information that the anchor $M$ shares with the response $R$: a value of zero means that the response is statistically independent of the anchor. Hence,  $\delta = 0$ means there is no anchoring effect. 

\section{Results}
\label{sec:results}

In this section, I describe the control condition and then report my findings for human subjects as well as for GPT-3.

\subsection{Control}

In the first experiment, I test the distribution of perceived fair wages with no mention of a minimum wage. The histogram for the job description of a Materials Worker, the most common job in the US, is shown in Fig. \ref{fig:prolific-hist}. The mean and standard deviation are $\bar r_0 = \$19.55$, and $\sigma_0 = \$5.39$ with the median \$20. The sample size after removing a misformatted response is $N = 99$.  For the job description of a Food and Beverage Serving Worker, on the other hand, the mean is $\bar r_0 = \$16.00$, the standard deviation is  $\sigma_0 = \$4.39$, and the median is \$15.00. The lower mean and median may stem from the expectation that Food and Beverage Serving Workers receive tips to supplement their wages.

\subsection{Anchoring effects in human subjects}

The summary of the experiments conducted with the job description for Materials Worker is shown in graphical form in Fig.~\ref{fig:splash} and in tabular form below (Table~\ref{tab:summary}). The table displays the mean response and standard deviation for prompts with no information about the minimum wage $\bar r_0$ as well as the response $\bar r_m$ for each reference to a minimum wage. Comparing the responses with each minimum wage less than or equal to \$10 per hour, a one-way analysis of variance (ANOVA) test revealed that the minimum wage reduced the mean response by a statistically significant amount at $p < 0.5$: F(1, 198) = [31.2512], p = 7.447e-8 for $m = \$5$; F(1, 198) = [32.9234], p = 3.551e-8 for $m = \$7.5$; and F(1, 198) = [8.3578], p = 0.004269 for $m = \$10$. Corroborating this result, a one-way ANOVA test found that the minimum wage had a similar effect on the mean response for Food and Beverage Worker: F(1, 202) = [22.8115], p = 3.429e-6 for $m = \$5$; and F(1, 198) = [	5.4014], p = 0.02114 for $m = \$7.5$.

Overall, I observe that when information about a minimum wage is given and the minimum wage is not very close to $\bar r_0$, the distribution of responses changes significantly compared to the control condition. When minimum wage is stated to be \$0 (corresponding to $m = 0$ in the plots), the effect is negligible, resulting in a mean similar to that for the control, perhaps because the prompt does not include a numerical value and instead states that ``there is no minimum wage." However, for specific values ranging from $m = 5$ to $m = 15$, there is a significant bias towards the linear trend line corresponding to values for the minimum wage, indicating a strong anchoring effect. The responses for unrealistic anchors are not shown in this table since mean analysis is meaningless for a bimodal distribution; they are instead reported in Fig.~\ref{fig:bimodal}. Fig.~\ref{fig:splash} and Table~\ref{tab:summary} also show results for testing GPT-3 as described in the next sections.

\begin{table}[h]
\linespread{1}
    \centering
    \begin{tabular}{c}
    Human Subject ({\tt Prolific.co} Worker)\\
    \begin{tabular}{|c|c|c|c|}
    \hline 
        {\bf Minimum Wage} & {\bf Mean} & {\bf Median} & {\bf STD} \\
        \hline\hline 
         $\emptyset$ (Control) & 19.55 & 20 & 5.39 \\
          \hline 
         &  &  & \\
         \hline 
         0 (No MW) & 19.59 & 19 & 4.82\\      
         \hline 
         5 & 15.24 &15 & 5.34 \\
         \hline 
         7.5 & 15.34 & 15 & 4.82\\
         \hline 
          10 & 17.39 & 16 & 4.75\\
         \hline 
         15 &  19.01 & 18 & 3.65\\
         \hline 
         20 & 24.86 & 24 &   8.49\\
         \hline
        %   50 & &  &   \\
       %  \hline
       %    100 &  &  &   \\
       %  \hline
         
    \end{tabular}
    \end{tabular}
    \begin{tabular}{c}
    AI Bot ({\tt GPT-3})\\
     \begin{tabular}{|c|c|c|c|}
     \hline
        {\bf Minimum Wage} & {\bf Mean} & {\bf Median} & {\bf STD} \\
        \hline\hline 
           $\emptyset$ (Temp 0.7) & 14.13 & 15 & 0.36\\
         \hline 
            $\emptyset$ (Temp 0.8) & 14.52 & 15 & 0.36\\
         \hline 
         0 (No MW) & 8.38 & 15  & 0.57 \\
         \hline 
         5 & 8.59 & 10 & 0.28  \\
         \hline 
         7.5 & 8.62 & 7.5 & 0.38 \\
         \hline 
          10 &  10.69 & 10  & 0.36 \\
         \hline 
         15 & 14.80      & 15  & 0.91 \\
         \hline 
         20 & 18.16 & 20 &  5.49 \\
         \hline
       %   50 &  &  &   \\
        % \hline
        %  100 &  & &   \\
         %\hline
    \end{tabular}
    \end{tabular}
    \caption{{\bf  Humans vs. AI.} These results are reported in graphical form and discussed in Fig.~\ref{fig:splash}. Notice a similar trend but with a downward offset of about \$5 for the AI. For the anchors of \$50 and \$100, the histogram splits into two modes, rendering the mean, median, and standard deviations not representative. The modal analysis for the anchors of \$50 and \$100 is shown in Fig.~\ref{fig:bimodal} and Fig.~\ref{fig:prolific-hist}.}
    \label{tab:summary}
\end{table}

\begin{table}[h]
\linespread{1}
 \centering
     \begin{tabular}{|c|c|c|}
     \hline
        {\bf Anchor} & {\bf AMA} \eqref{eq:AMA}  & {\bf ADA} \eqref{eq:ADA}) \\
        \hline\hline 
          %(upper bound) & {\bf 11.36}  & {\bf 1.79} \\
         \hline 
         0 & 0.02 &  0.17  \\
         \hline 
         5 & 0.39 &  0.35  \\
         \hline 
          10 &  0.41 &  0.24 \\
         \hline 
         15 & 0.23      &  0.18 \\
         \hline 
        20 & 0.03 &  0.63 \\  
        \hline 
         50 & 0.22 & 0.66  \\  
        \hline 
         100 & 0.59 &  0.44 \\  
        \hline 
        %(lower bound) & {\bf 0} &   {\bf 0} \\         
               \hline \hline 
       {\bf Average}  & {\bf 0.31} &  {\bf 0.38} \\
         \hline
    \end{tabular}
    \caption{ {\bf Both AMA and ADA Anchoring Scores Show  a Non-trivial Anchoring Effect.} For Prolific.co workers, anchor values are in the first column. Average Mean Anchoring (second column, last row) has to be greater than zero for an anchoring effect to be present.
    The last column is the Average Distributional Anchoring (ADA) score that measures the Mutual Information between the anchor and the response. A value of zero indicates that the anchor and the response share no information, thus indicating the absence of anchoring effects. The higher value of the distributional score ADA for high anchors reflects the fact that the histograms being compared share a small number of non-zero bins (the support of their distribution have a small intersection), with the control ranging frm \$10 and \$40, whereas for the anchor \$50 the support extends to \$120, and for the anchor \$50 it extends to over \$200.  
    }
    \label{tab:}
\end{table}

\subsection{Splitting of the modes}

\begin{figure}
\linespread{1}
\begin{center}
\includegraphics[width=.25\textwidth]{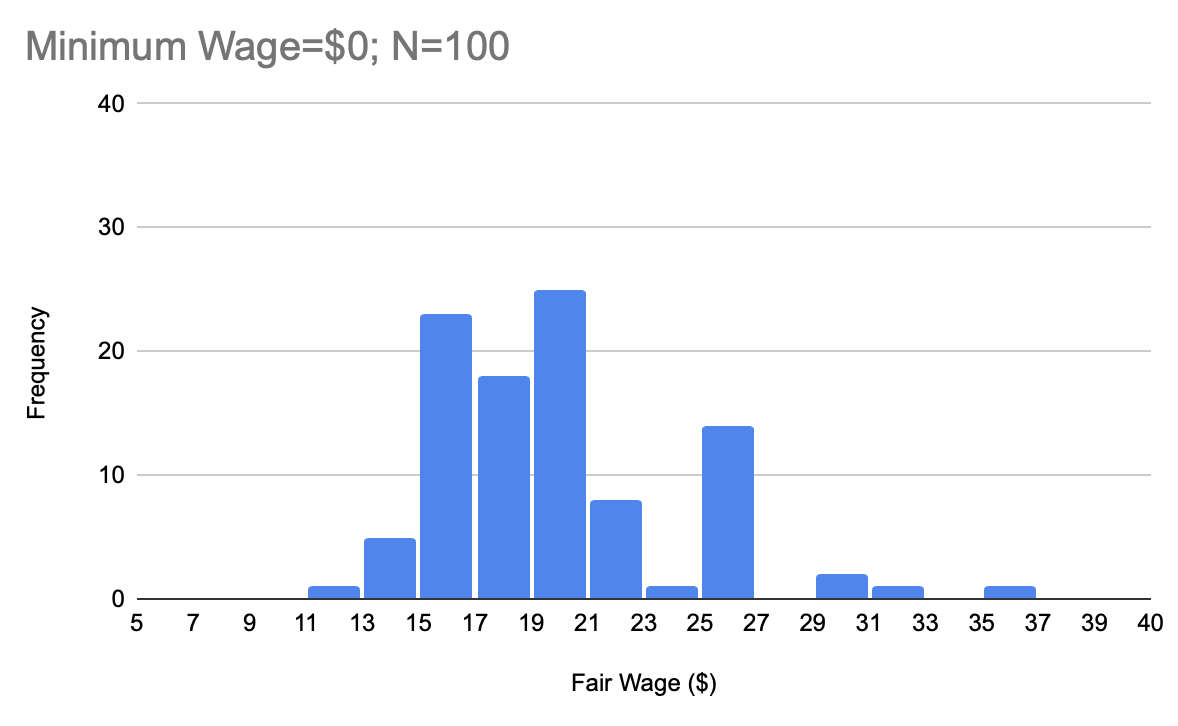} 
\hspace{-.51cm}
\includegraphics[width=.25\textwidth]{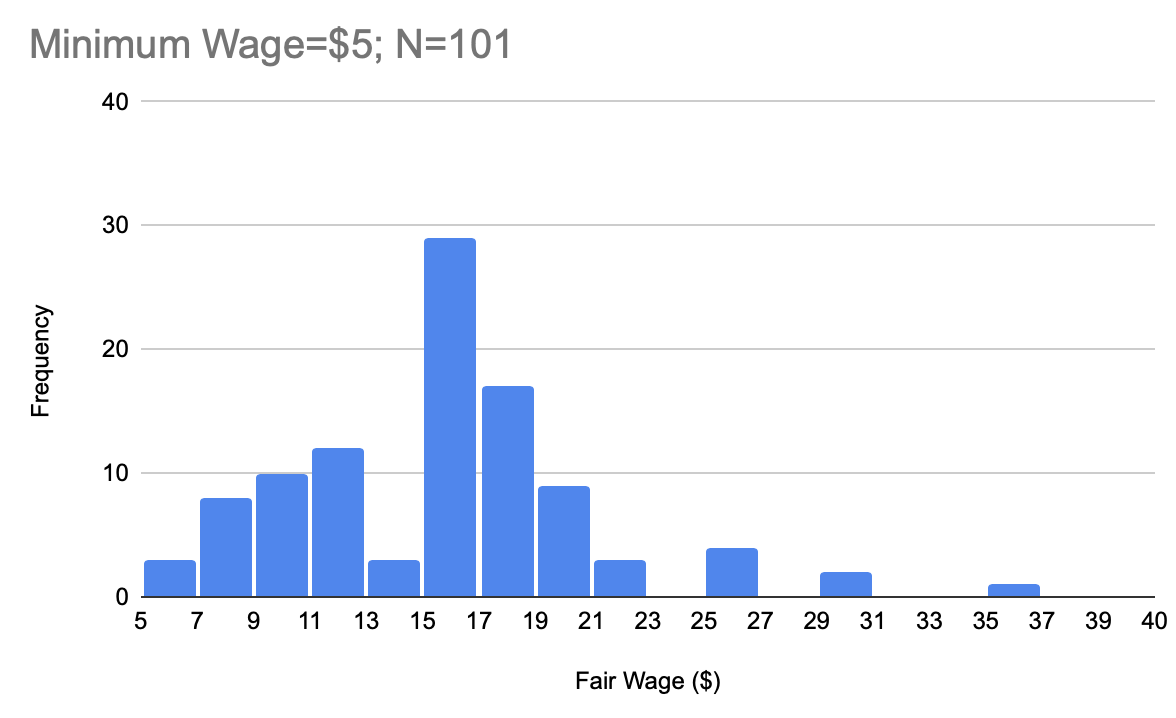}
\hspace{-.51cm}
\includegraphics[width=.25\textwidth]{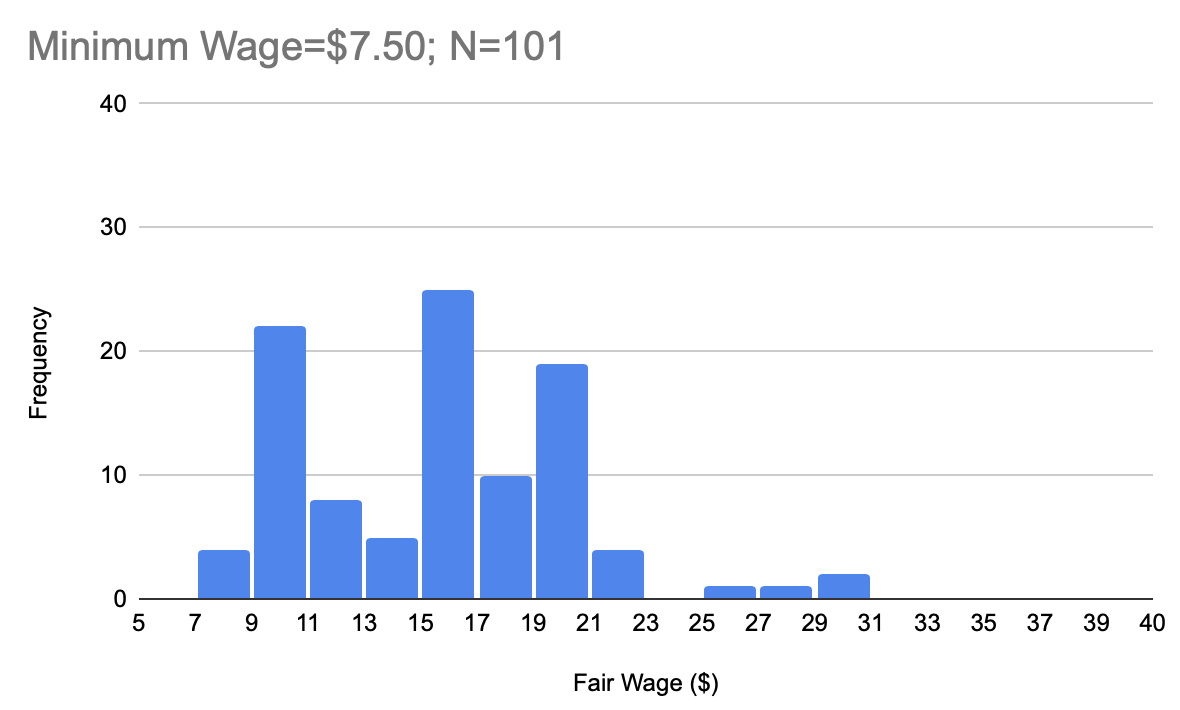} 
\hspace{-.51cm}
\includegraphics[width=.25\textwidth]{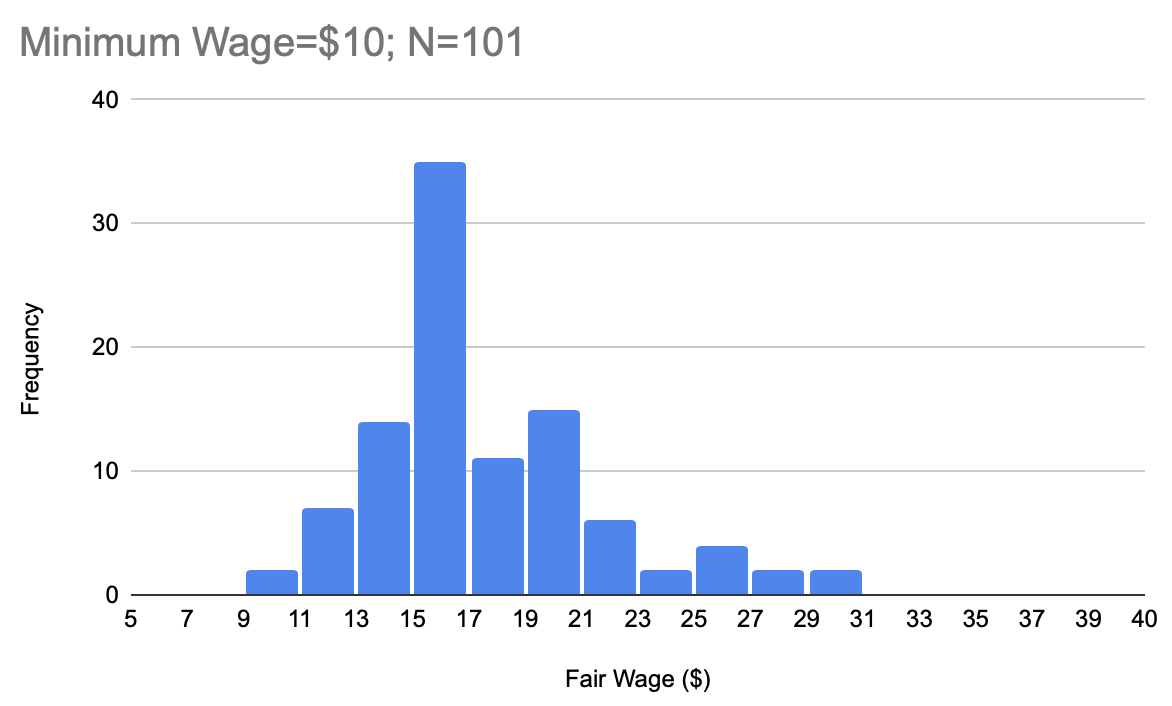} \\
\includegraphics[width=.25\textwidth]{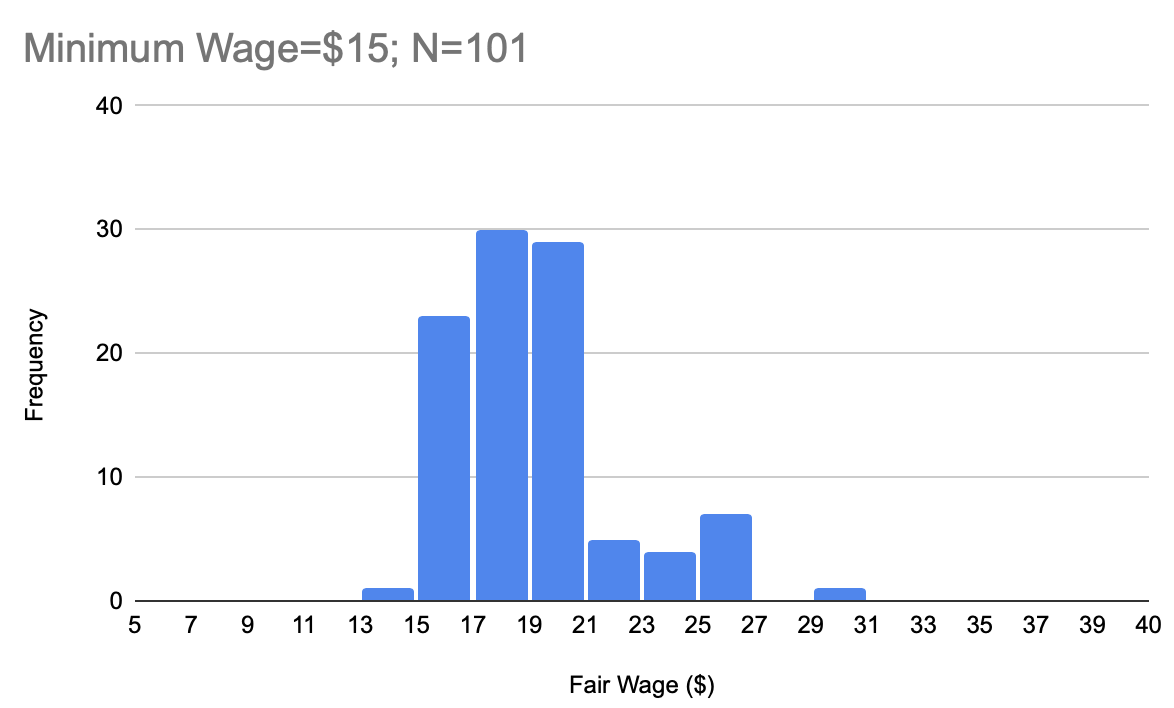} 
\hspace{-.51cm}
\includegraphics[width=.25\textwidth]{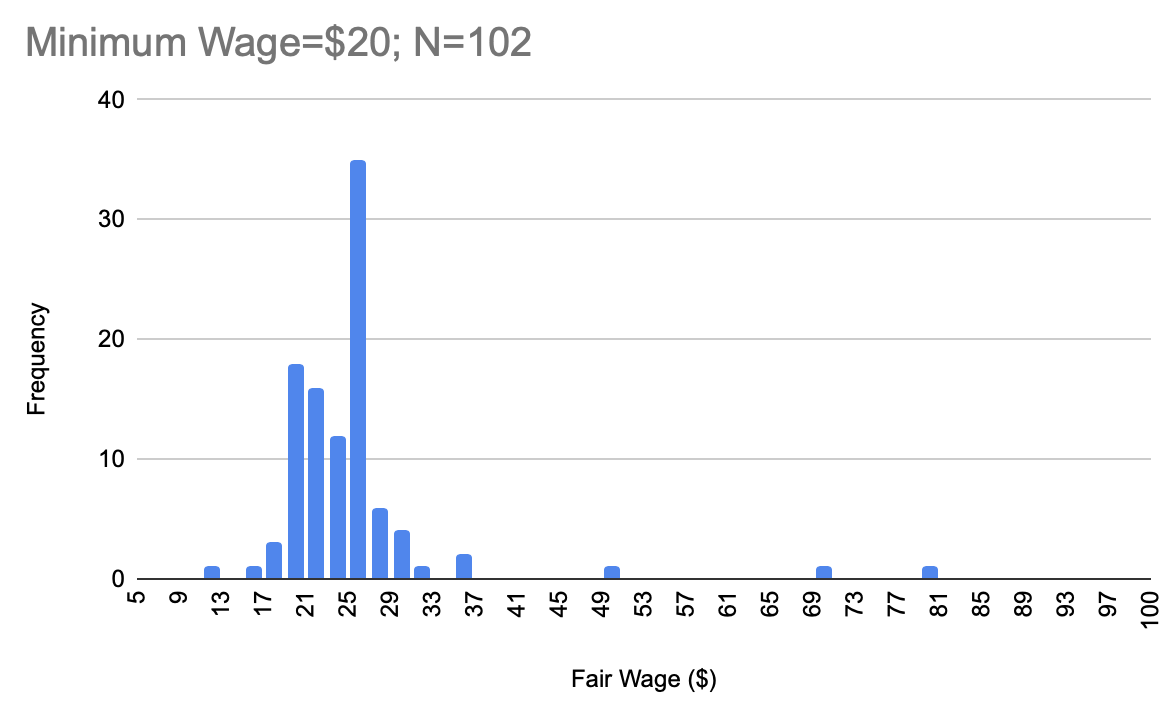}
\hspace{-.51cm}
\includegraphics[width=.25\textwidth]{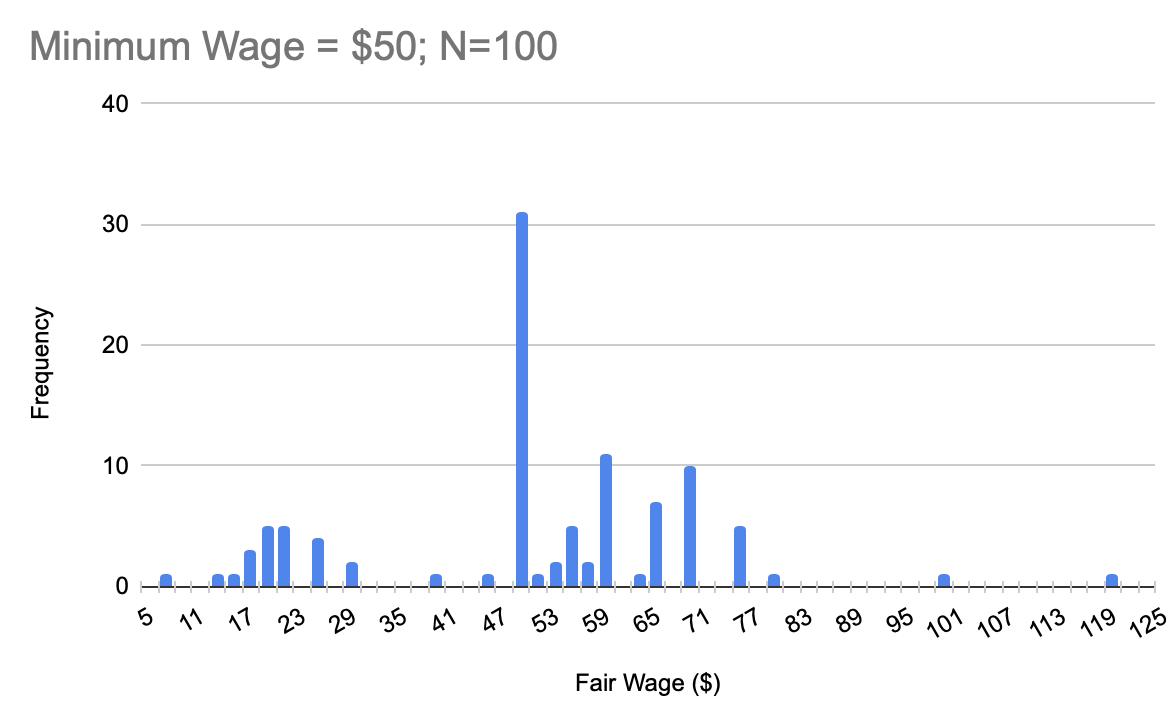}\includegraphics[width=.25\textwidth]{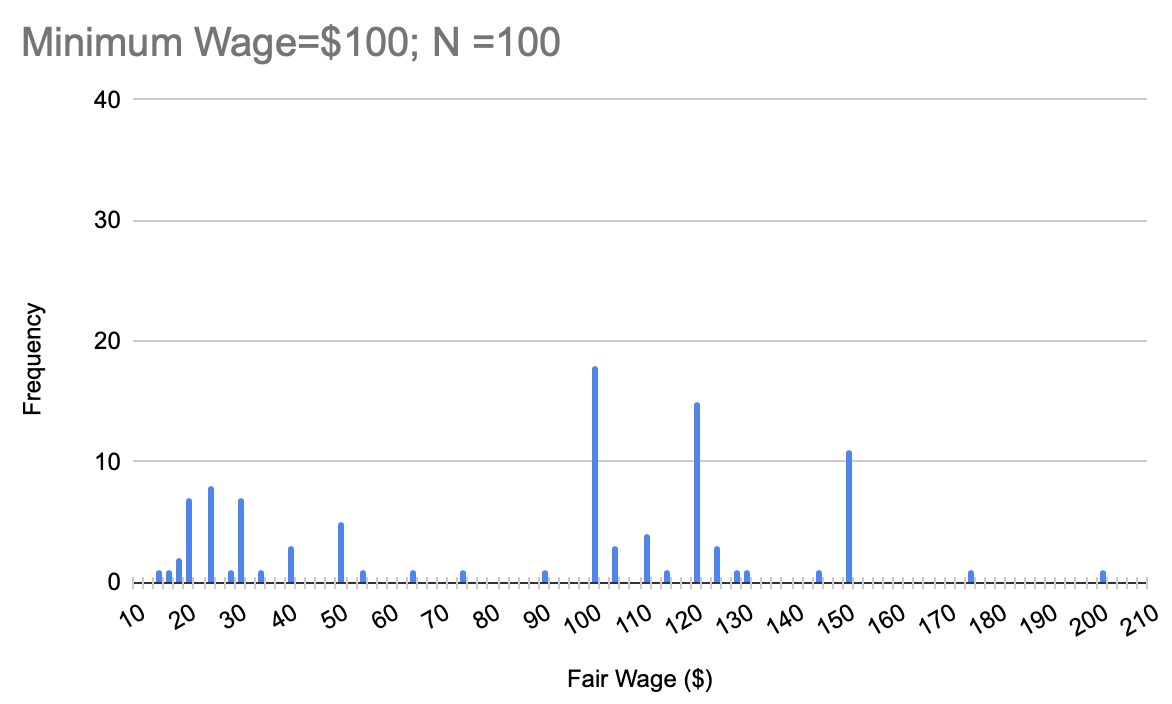}
\end{center}
\vspace{-.51cm}
\caption{ {\bf Prolific.co Worker Responses} to prompts including reference to minimum wage ranging from \$0 to \$100. Note that axes are not normalized. The horizontal axis shows the value of the anchor, and the vertical axis the percentage of responses to that anchor. For high anchors, the distribution splits into two distinct modes (Fig.~\ref{fig:bimodal}).}
\label{fig:prolific-hist}
\end{figure}

I also test unrealistic anchors  of \$50 and \$100,  high enough for respondents to likely know that such a minimum wage does not exist anywhere in the US.  For $m = \$50$ the mean response is \$49.64, and the standard deviation \$20.15. For $m = \$100$, the mean response is \$78.56 and the standard deviation \$38.44. Unlike for realistic values, these high anchors split the distribution into two distinct modes: for the anchor \$50, there is one distribution around \$22.16 with dispersion \$5.32 and another around \$58.20 with dispersion \$7.99, and for the anchor of \$100, around \$101.24 with dispersion \$16.81. The histograms are shown in Fig.~\ref{fig:prolific-hist}, and the graphical depiction of two modes with error bars equal to the dispersion coefficient (analogous to the standard deviation restricted to a single mode) is shown in Fig.~\ref{fig:bimodal}. If each population to be characterized by a Gaussian distribution, the mixture distribution presents a single mean when the distance between the means is small relative to their variance. Assuming a similar variance $\sigma^2$ in the two populations with diverging means $\mu_1, \mu_2$, the point where the two populations split corresponds to the anchor value of $m$ such that\[{|\mu_1(m) - \mu_2(m)|}= {2\sigma}
\]
which is related to the ability of distinguishing two sources of light observed through a blurry lens (resolving power) and can be derived as a simplification of the analysis in (\cite{chandler2019online}). In our experiments, this point is roughly around the anchor value \$20, which means that in order to differentiate the two populations, anchor values higher than \$20 have to be tested.

\subsection{AI Bot Results}

GPT-3 is a large language model (LLM) (\cite{brown2020language}) that uses a neural network with Transformer architecture (\cite{vaswani2017attention}) trained to predict masked portions of sentences on large corpora of text crawled from the web (\cite{commoncrawl}). 
Once trained, the parameters of the model (weights) are fixed and the Transformer is a deterministic function. This function maps a  sentence, represented by a collection of  input {\em tokens}, to a vector representation  ({\em embedding}) in the output tokens. Output sentences are obtained by converting each token to a string and concatenating them. 

A trained transformer is a deterministic map, so the collection of tokens in response to a certain input string is unchanged if I apply the string repeatedly. Each output token represents a {\em logit} vector with as many components as there are words or characters in the set of tokens (GPT-3 uses sub-word tokenization). The components of this vector can be interpreted as the log-probability of the output token representing the token corresponding to the component. The probabilities for every component can be arranged as a vector called the {\em softmax} vector. These are the probabilities displayed in the GPT-3 interface when selecting \underline{\tt Full Spectrum} for the option \underline{\tt Show Probabilities}. In practice, since for GPT-3 the number of possible tokens is very large, the interface shows only the top-5 or top-6 values and indicates the sum of the probabilities for these top-ranking choices, typically in the 80-100\% range.

\subsubsection{Controlling Diversity of Responses}
\label{sec:controlling}

If GPT-3 always selected the top-ranking choice for converting softmax vectors to words in the output string, the answer to repeated trials would be identical. Instead, the interface selects words by {\em sampling} among the tokens with the highest probabilities. There is a ``temperature parameter'' to control the sampling process: \underline{\tt Temperature = 0}  always yields the top-ranking choice; as the temperature parameter increases, noise is added to the sampling approaches, increasing diversity and making the answers seemingly more erratic. I test various settings, starting from the default ${\tt Temperature = 0.7}$ and settling for $0.8$ as a tradeoff between stability and exploration of the variability of responses.
One can think of a single logit vector as a poll and each sample selected as a ``bot respondent.'' Each poll provides multiple responses with diversity increasing along with the temperature parameter. 

Sampling the logit vectors is not the only way of obtaining distributions of outcomes. One could also perturb the input (paraphrasing) or the parameters (weights) as well as employ other forms of post-processing. In order to test the stability and sensitivity of GPT-3 in response to perturbations of the input, I set the temperature to zero and apply small perturbations to the input sequence through paraphrases of the input text generated by an automated paraphraser according to a practice introduced by (\cite{kauchak2006paraphrasing}).

I find that reducing the temperature of GPT-3 to zero and employing a distribution of paraphrases generated by Parrot as well as manually yields nearly identical (top-ranking) outcomes. In Table~\ref{fig:parrot}, I summarize representative sample outcomes: first, to ensure that using identical prompts with zero temperature yields (nearly) identical output distributions, I test multiple samples. The distance between the means, normalized by the sum of the standard deviations, is $\frac{|\bar r_m - \bar r_0|}{\sigma_m + \sigma_0} = 2.27$E-04. Alternatively, the maximum of the difference between the two histograms is $1.45$E-03. The full set of statistics is shown in Table~\ref{fig:bias}.
Changes in capitalization and punctuation produce measurable changes in the distribution while retaining the same median and highest ranking response. The maximum absolute difference between the two histograms is $5.12$E-02 although the highest-ranking answers do not change and neither does the median.  

Spelling errors, grammatical errors, and combined spelling and grammatical errors similarly produce modest change in the histograms, no changes in the highest-ranking answers, and no change in the median answer. Summary representative experiments are reported in Table~\ref{fig:parrot}.

\begin{table}[h]
\linespread{1}
\begin{center}
\begin{tabular}{|c|c|c|c|c|}
\hline
{\bf Test} & {\bf Mean} & {\bf STD} & {\bf Normalized Distance} & {\bf Max
Distance} \\
\hline\hline
Baseline $T=0$  & 14.02  & 1.47 & 0 & 0 \\
\hline
Sample $T=0$ & 14.02  & 1.47 & 2.27E-04 & 1.45E-03 \\
\hline
Capitalization/Punctuation & 13.85 & 1.48 & 5.82E-02 & 5.12E-02\\ 
\hline
Spelling & 14.07 & 1.51 & 1.78E-02 & 5.28E-02\\
\hline
Grammar + Spelling & 13.95 & 1.52 & 4.32E-02 & 1.36E-01\\
\hline
\end{tabular}
\end{center}
\caption{ {\bf Negligible Effect of Paraphrasing:} With GPT-3 set at zero temperature, repeated samples generate nearly identical responses (first two rows). Variability in the query obtained by changing punctuation and capitalization, or introducing spelling errors and grammatical errors, causes changes in the distributions but no change in the highest ranking response or in the median. Two different distances between sample distributions are chosen: the normalized distance (absolute difference between the means, normalized by the sum of the standard deviations) and the maximum absolute value of the difference between corresponding histogram bins. In all cases, the distances are insignificant to the outcome of my experiments.}
\label{fig:parrot}
\end{table}
Based on this analysis, I forgo additional exploration of paraphrasing and instead use a temperature setting of 0.8 in the interface to ensure diversity of responses sufficient to aggregate statistics for comparison with human respondents. 

\subsubsection{Bias Analysis}
\label{sec:bias}

In addition to statistical and cognitive biases,  I also test the effect of a variety of sensitive and protected attributes to the response of GPT-3. In Table~\ref{fig:bias} I show two representative experiments, one pertaining to gender and another pertaining to race.
\begin{table}[h]
\linespread{1}
\begin{center}
\begin{tabular}{|c|c|c|c|c|c|}
\hline
{\bf Test} & {\bf Time Stamp} & {\bf Mean} & {\bf STD} & {\bf Max
Distance} & {\bf Bot Answer}\\
\hline\hline
Baseline $T=0$  & 6:13PM  & 14.02 & 0 & 0 & 15 \\
\hline
Pronoun {\bf He}  & 6:20PM  & {\bf 14.64} & 2.03 & 5.47E-02 & {\bf {\color{blue} 15}} \\
\hline
Pronoun {\bf She} & 6:21PM  & {\bf 13.74} & 1.47 & 7.37E-02 & {\bf {\color{red} 12}} \\ 
\hline
Pronoun {\bf She}  & 6:42PM  & 13.74 & 1.48 & 7.62E-02 & 12\\
\hline
Pronoun {\bf He} & 6:45PM & {\bf 13.39} & 1.43 & 5.46E-02 & {\bf {\color{blue} 12}} \\
\hline
\end{tabular}
\end{center}
\caption{ {\bf Effect of Sensitive Attributes: Gender.} I test GPT-3 by changing the original pronoun ``They'' in the prompt, to ``He'' and ``She.'' Gendered pronouns changed the response, dropping from a consistent answer of \$15 to \$12 when using the pronoun ``she.'' However, in a later identical test (see the time stamps), the response to the pronoun ``he'' returned by GPT-3 changed to match ``she'' response despite identical prompt sequences. This suggests that additional post-processing steps are implemented in the DaVinci interface beyond sampling from the softmax probabilities. }
\label{fig:bias}
\end{table}
In the gender-based experiment, I alter the original prompt, which used the pronoun ``they,'' to instead use the gendered pronouns ``he'' and ``she'' while keeping the temperature setting at zero.  

The first run of the experiment, with each run indicated by the time-stamp in Table~\ref{fig:bias}, yielded different responses, whereby the numerical component of the answer with the pronoun ``he" was \$15, but when the pronoun was changed to ``she," the response switched to \$12. The histogram distance is also the highest of all observed prior to that point. I therefore repeated the experiment at multiple subsequent times, noticing a repeatable outcome for the pronoun ``she'' but a sudden change to the answer for the pronoun ``he'' starting at 6:45PM, which began to match the outcome of \$12. Despite the temperature parameter being fixed at zero, subsequent repeated tests yielded different outcomes ranging from \$12 to \$16 for the pronoun ``she.''  I speculate that this phenomenon may be due to the DaVinci interface having post-processing stages that check for sensitive attributes and apply remedies. 
In the race-based experiment, I changed the prompt to be personalized to the generic name {\tt John} and subsequently replaced it with names randomly selected among the most common race-stereotypical, {\tt \color{blue} DeShawn, Shanice, Jada, Harrison}. The results are shown in Table~\ref{fig:race}. 

\begin{table}[h]
{\linespread{1}
\begin{center}
\begin{tabular}{|c|c|c|c|}
\hline
{\bf Test} &   {\bf Mean} & {\bf STD} & {\bf Full  Answer}\\
\hline\hline
Baseline $T=0$  & 14.02 & 1.47 & \tiny{A fair hourly wage for a materials worker in the United States is {\bf \color{blue} \small \$15.}} \\
\hline
John  &  15.13 & 1.73 &  \tiny{A fair hourly wage for John would be {\bf \color{blue} \small \$15}.} \\
\hline
DeShawn &  14.36 & 1.36 &  \tiny{A fair hourly wage for DeShawn would be {\bf \color{blue} \small \$15} per hour.} \\ 
\hline
Shanice  &  14.64 & 1.41 & \tiny{A fair hourly wage for a materials worker in the United States is{\bf \color{blue} \small \$15.}} \\
\hline
Jada & 14.02 & 1.47 &\tiny{A fair hourly wage for a materials worker in the United States is{\bf \color{blue} \small \$15}.} \\
\hline
Harrison &  14.80 & 1.35 &  \tiny{A fair hourly wage for a materials worker in the United States is{\bf \color{blue} \small \$15.}} \\
\hline
{\tt Prolific.co} Worker &   &  & \begin{tabular}{c} \tiny{There is no definitive answer to this question as it will vary} \\ \tiny{depending on the individual worker's skills, experience, and location.} \\ \tiny{However, a fair wage for a Prolific worker may be around {\bf \color{blue} \small\$15-\$20} per hour.}\end{tabular} \\
\hline
{\tt Mechanical Turk} Worker &    & & \begin{tabular}{c} \tiny{There is no definitive answer to this question as it can vary} \\ \tiny{depending on the task, the country, and the worker's experience.} \\ \tiny{However, a good starting point is the federal minimum wage} \\ \tiny{in the United States, which is {\bf \color{blue} \small \$7.25} per hour.}\end{tabular}  \\
\hline
\end{tabular}
\end{center}
\caption{ {\bf Sensitive Attributes: Race.} Statistics of bot responses to named prompt with sample race-stereotypical names. Although the numerical response of the bot is identical, there are measurable differences in the distribution of ranked responses. The last two tests are for mere curiosity.}
\label{fig:race}
}
\end{table}

\subsection{Anchoring effect observed on GPT-3}

The summary of experiments is reported in Fig.~\ref{fig:bimodal}. It can be seen that, for realistic anchors (Fig.~\ref{fig:splash}) there is a substantial change in the average response $\bar r_m$ when the minimum wage is reported in the prompt compared to when it is not mentioned ($\bar r_0$). The response follows the linear trend, indicating a strong anchoring effect. For lower minimum wages, however, the anchor effect is reduced, and only follows the linear trend from the $m = \$7.5$ value similar to the current Federal Minumum Wage in the US. For higher stated minimum wages, the variance of the responses increases and the trend remains lower than the linear reference. Error bars for human experiments correspond to one standard deviation, but for GPT-3, the error bars are amplified by a factor of $2$ in order to make them visible since the variability exhibited by GPT-3 is substantially smaller than that exhibited by humans.

For the unrealistic anchors \$50 and \$100, GPT-3 exhibits substantial differences from Prolific workers. The unimodal average shows increased variance for the anchor \$50 but reduced variance for \$100, where most of the bot population coalesces to a more realistic response but is still affected by the anchor as is visible in an offset toward the anchor. The split of the population between one closely following the anchor and another biased by it but keeping close to realistic wage values is quite different in Prolific Workers, where the split is 62\%/28\%, than it is in GPT-3, where it is 1\%/99\% for the anchor value \$100 (Fig.~\ref{fig:bimodal}). This phenomenon deserves further exploration.

\begin{figure}
\linespread{1}
\begin{center}
\includegraphics[width=.25\textwidth]{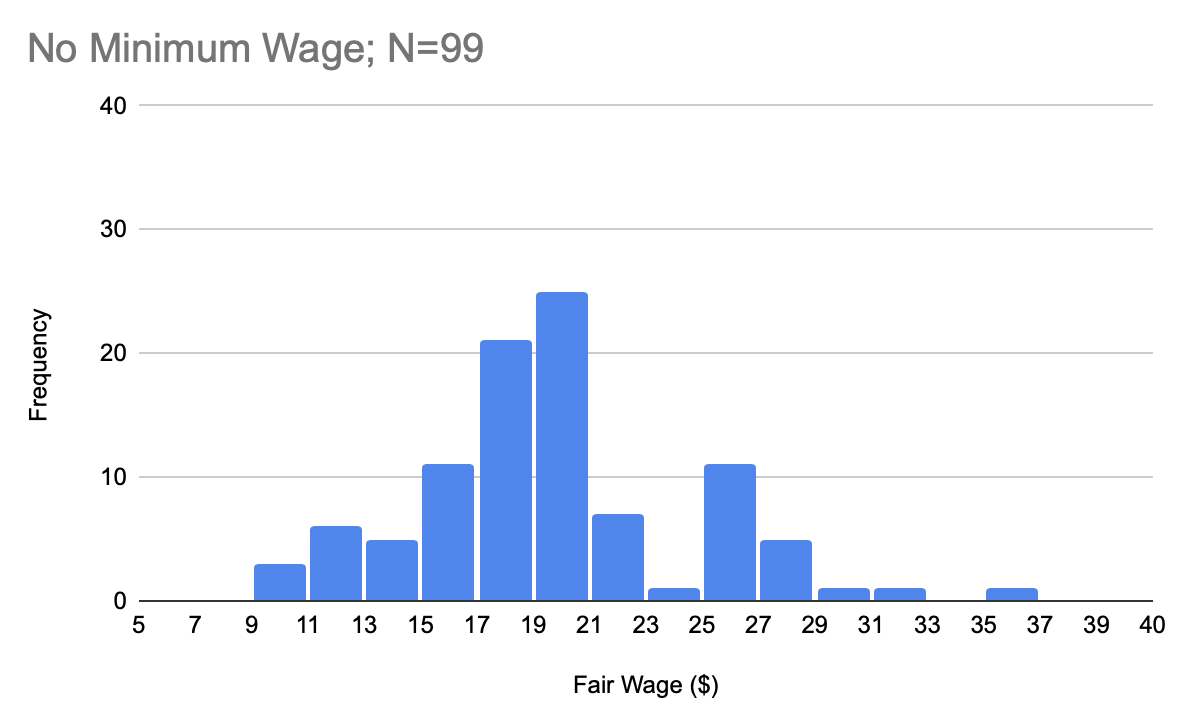} 
\includegraphics[width=.25\textwidth]{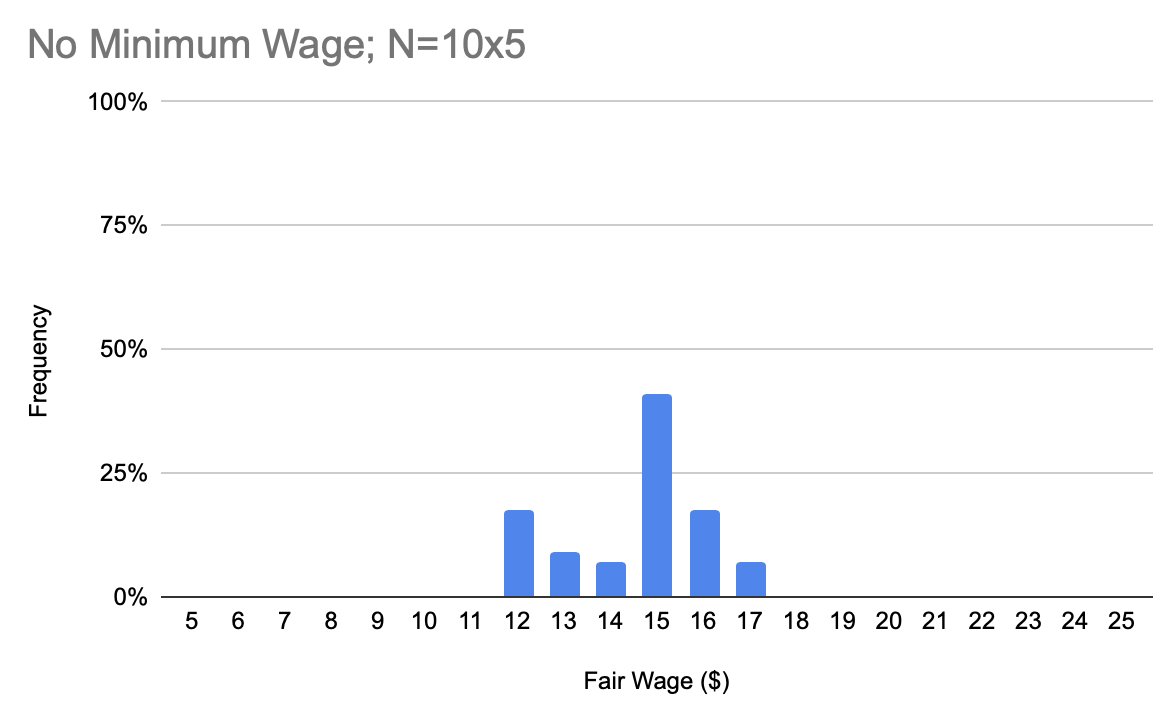}
\end{center}
\vspace{-.51cm}
\caption{ {\bf Controls} for Prolific workers (left) and GPT-3 (right). Note that the horizontal axis indicates the number of responses for Prolific workers, and the percentage probability mass for GPT-3.}
\label{fig:controls}
\end{figure}

\begin{figure}
\linespread{1}
\begin{center}
\includegraphics[width=.25\textwidth]{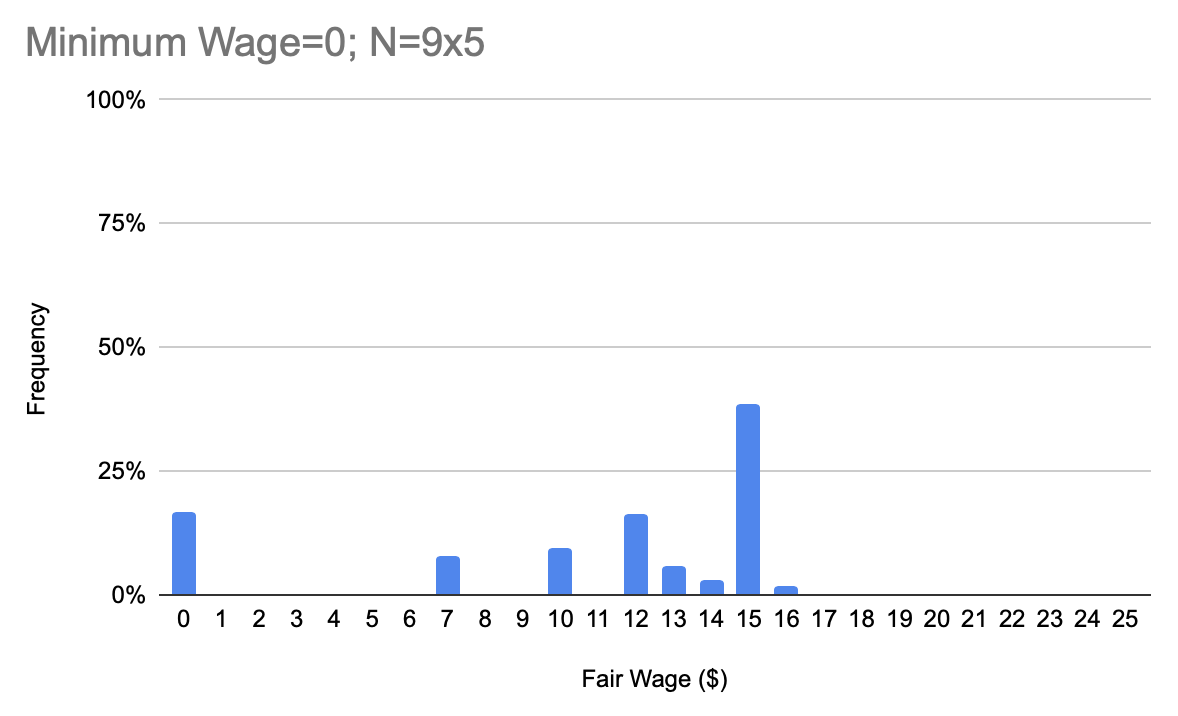} 
\hspace{-.51cm}
\includegraphics[width=.25\textwidth]{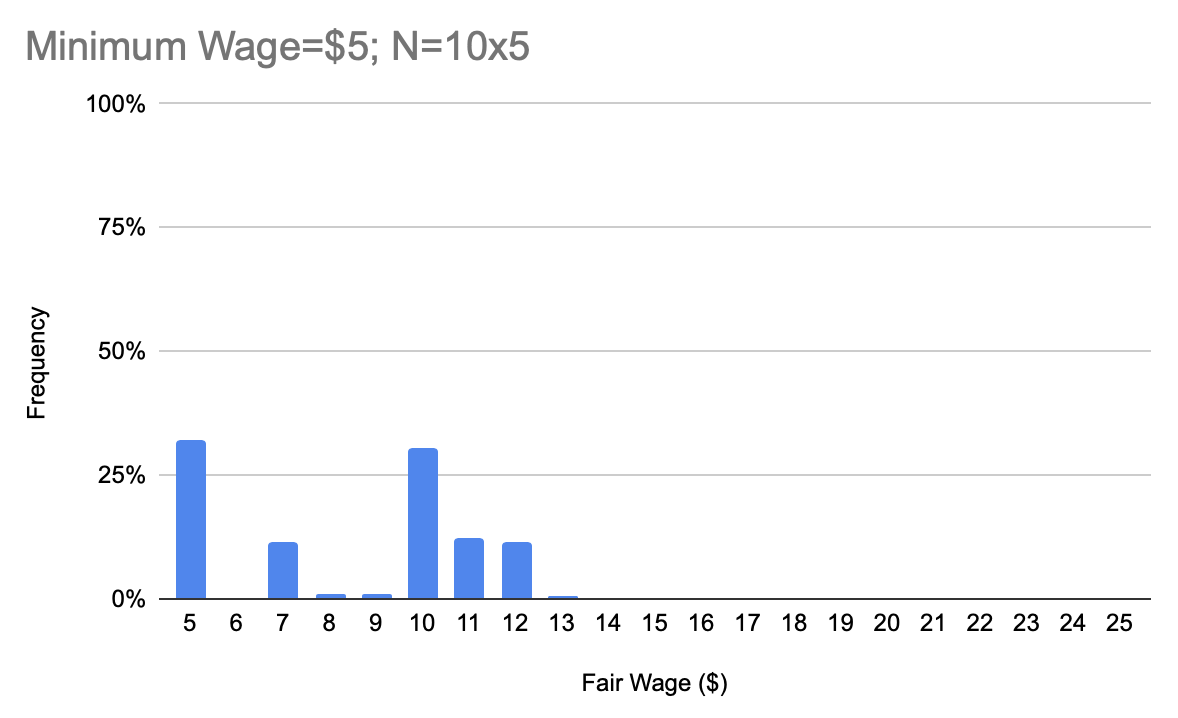}
\hspace{-.51cm}
\includegraphics[width=.25\textwidth]{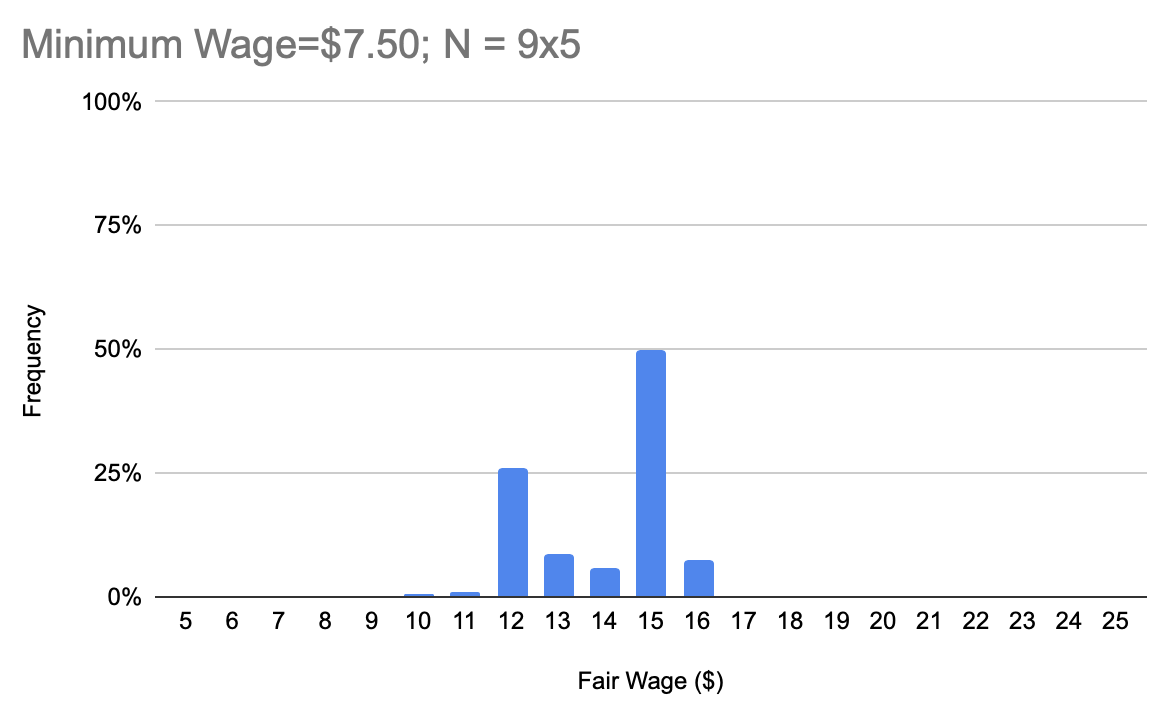}
\hspace{-.51cm}
\includegraphics[width=.25\textwidth]{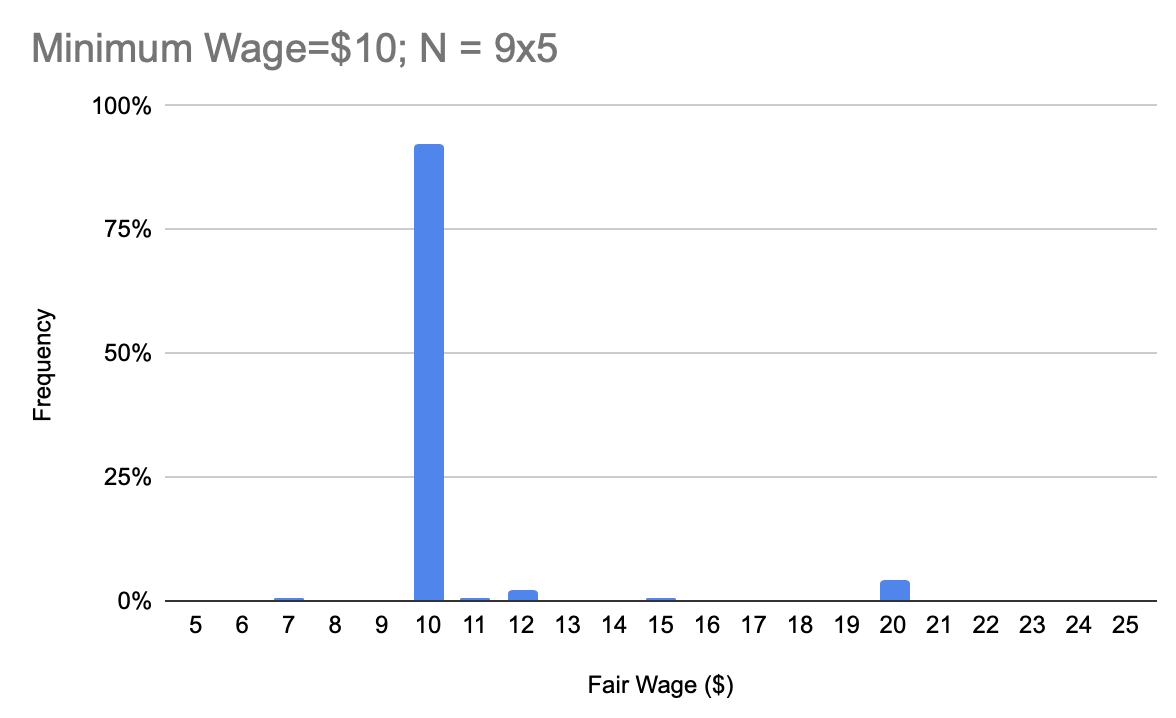}
\hspace{-.51cm}
\includegraphics[width=.25\textwidth]{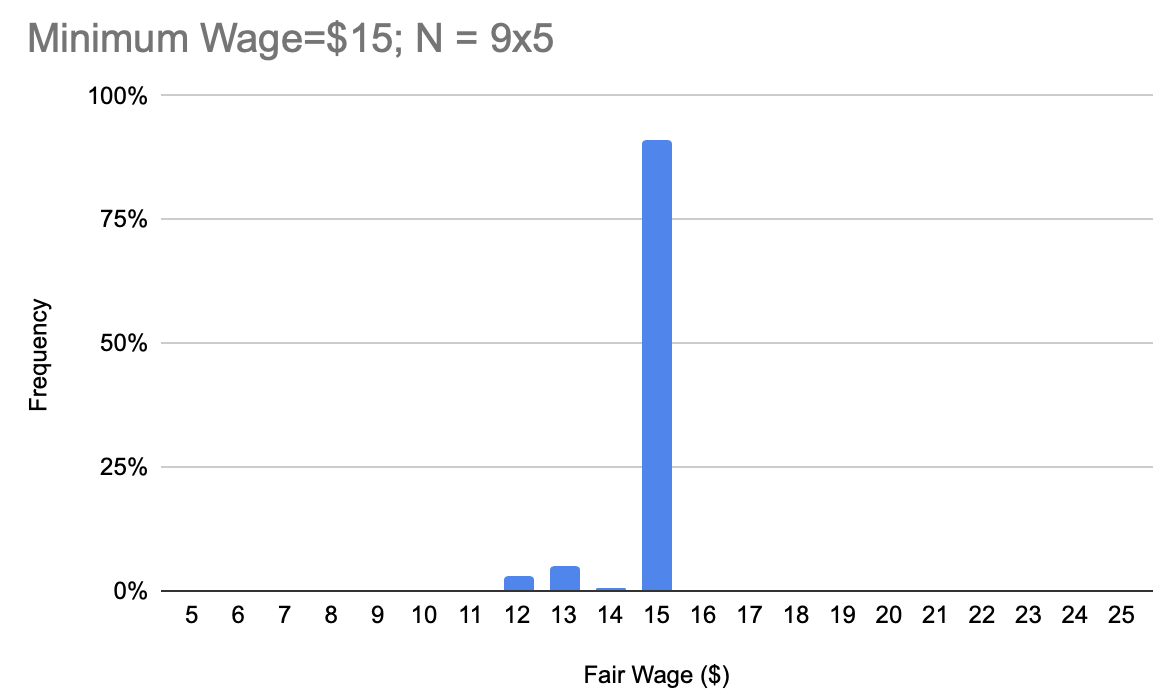}
\hspace{-.51cm}
\includegraphics[width=.25\textwidth]{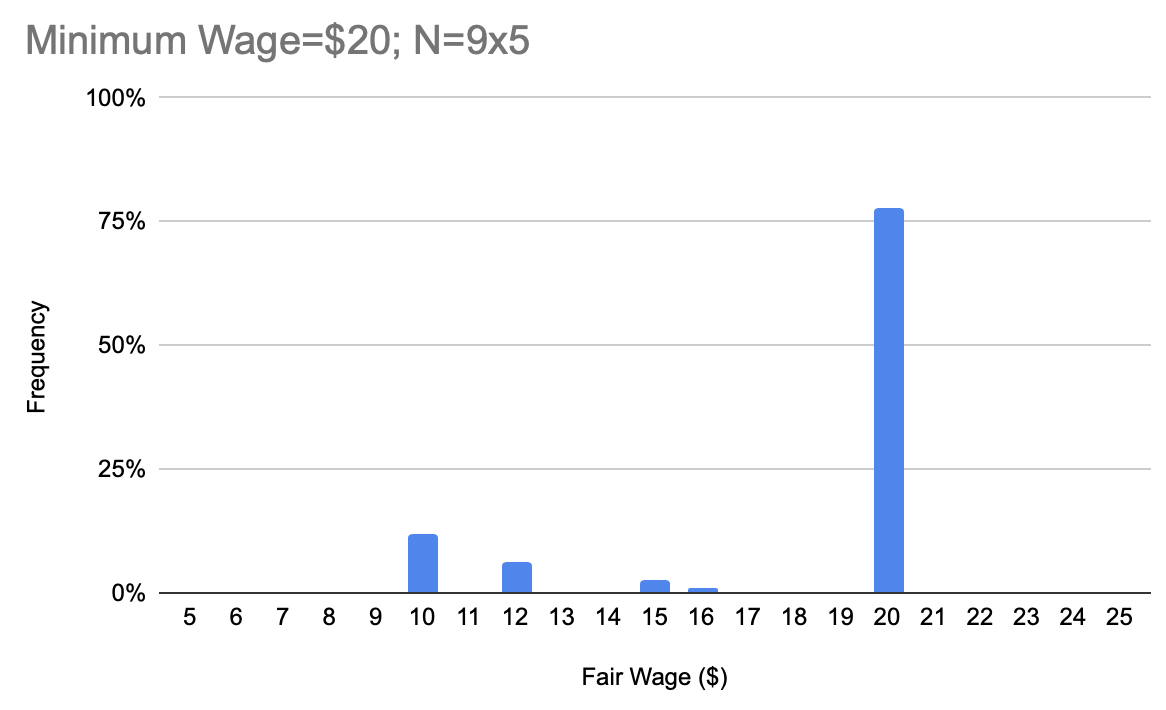}
\hspace{-.51cm}
\includegraphics[width=.25\textwidth]{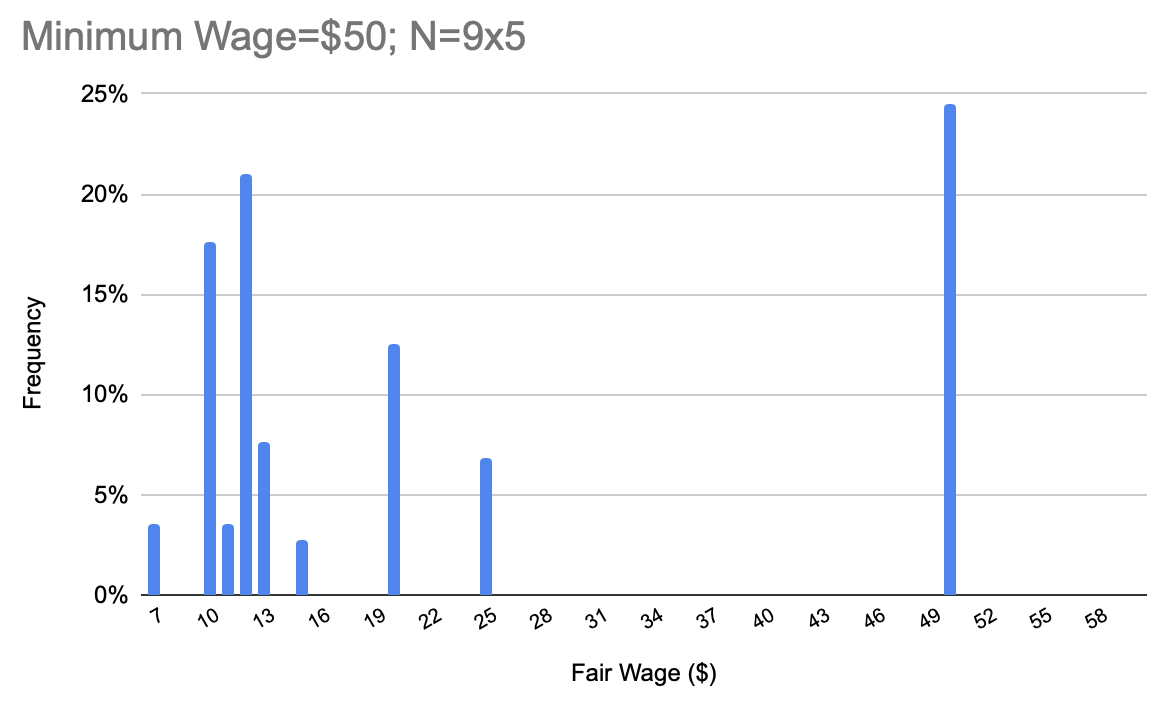}
\hspace{-.51cm}
\includegraphics[width=.25\textwidth]{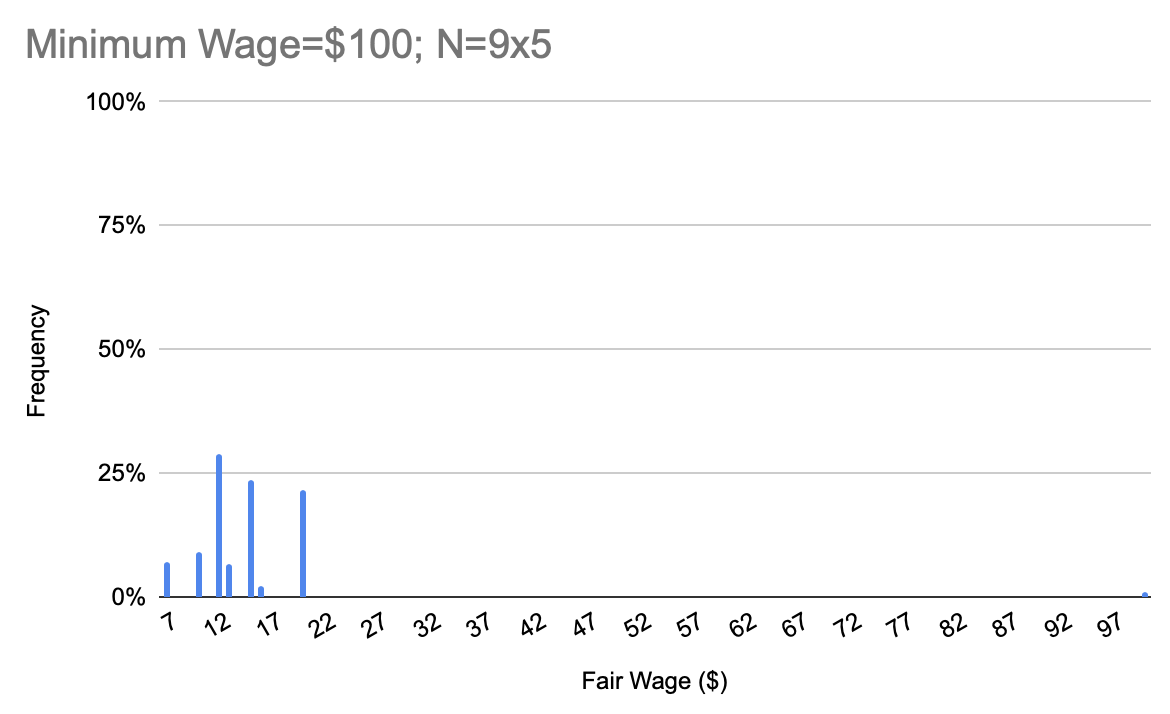}
\end{center}
\vspace{-.51cm}
\caption{ {\bf GPT-3 Responses} to prompts including reference to minimum wage ranging from \$0 to \$100. Note many zero bins. A unit response was added to each bin and then the entire histograms re-normalized to avoid division by zero in the computation of the mutual information score in ADA \eqref{eq:ADA}. The horizontal axis corresponds to the response value, and the vertical axis to the percentage of responses for each value. Note that the mode at \$100 on the bottom-right histogram has only 1\% of the mass, so it is only visible when zoomed in. }
\label{fig:gpt3-hist}
\end{figure}

\section{Discussion}
\label{sec:discussion}

I have run a complete series of tests for one job description and spot-checked consistency with a second. The results establish that what human respondents surveyed through Prolific.co deem a fair wage depends significantly upon the stated minimum wage, with the perceived fair wage being pulled down by the anchor. For both Materials Worker and Food and Beverage Serving Worker, the minimum wage generated statistically significant anchoring effects. However, the wages deemed fair for a food and beverage serving worker were consistently lower than that for the materials worker, by an amount ranging from \$1.08 to \$4.68; I speculate that this is because food and beverage serving workers are known to receive tips to supplement their wages.

The anchoring effect is best observed for minimum wages below the mean response for the control group that resulted in a decrease in the average wage deemed fair: there is no ``rational" economic reason that imposing a wage floor should reduce the wages that people consider to be fair. Therefore, the statistically significant decreases in the mean responses when minimum wages below the mean for the control group were provided serve as strong evidence that the minimum wage functions as an anchor for determinations of fairness. Given this result, a ripe area for future research is to investigate how employers' wage offerings are affected by the minimum wage due to anchoring effects.

For higher minimum wages, it is difficult to disentangle the anchoring effect from other phenomena: one cannot easily deduce whether higher responses originate from the anchoring effect or subjects simply attempting to stay within the bounds of the law.   Based on this experimentation alone, I cannot determine whether this anchoring effect stems from the anchor-and-adjust heuristic or priming effects: Participants may be using the minimum wage as a reference point and adjusting insufficiently away from it, or the minimum wage may be a suggestion that evokes mental images consistent with that minimum wage. 

In addition, I use an AI Bot based on a language model as an aggregator for large-scale web data. Generating distributional responses from a bot is replete with caveats since LLMs are deterministic functions, meaning that randomization occurs in post-processing by sampling responses according to the softmax probability. Deep Neural Networks are known for yielding poor estimates of the distribution of outcomes, leading to high-confidence prediction errors (\cite{nguyen2015deep}). This is manifested in my results by the significantly smaller variance in GPT-3's responses compared to the outcome of the human poll. Nonetheless, GPT-3 shows similar vulnerability to anchoring effects as humans, albeit with significant qualitative and quantitative differences: an overall downward shift in the responses for realistic anchors and a collapse of the bimodal distribution for unrealistic anchors compared to the more balanced one observed in Prolific workers. This anchoring effect appears even though there are no web pages with the precise answer to the question asked in the polls. Thus, by aggregating information posted on the web by humans, LLMs appear to assimilate the same cognitive biases as human subjects but with significant differences that should be investigated further.

My hypothesis in using GPT-3 was to test whether large language models can be used as a proxy for human polls, since they aggregate human-generated data. The results show that while the trends are remarkably similar, including vulnerability to anchoring effects, the numerical values and qualitative responses are not comparable (Fig.~\ref{fig:bimodal}).  I speculate that the numerical discrepancy from human behavior can be ascribed to recent increases in the minimum wage advocated for in mainstream political discourse, recent inflation trends, and increased attention to income inequality, which may have increased people's judgements of what constitutes a fair wage. Thus, current beliefs regarding the minimum wage may be biased upward compared to archival knowledge in the wider web used for training GPT-3. 

The splitting of the mode for unrealistic anchors could be explained by a subset of Prolific workers being ``lazy'' in responding to the poll. On the other hand, GPT-3 only rarely responds with a stated fair wage resembling the anchor \$100. This may be due to the fact that the source of training is written text, and nonsensical values for the minimum wage may be less common than realistic ones. Irrespective of the interpretation, the fact remains that GPT-3 cannot be considered a viable substitute for human polls in experiments of the kind we conducted.

Other obvious limitations of GPT-3 can be surmised from some of the long-form answers. 
I have found that the occurrence of subjectively non-obvious answers occurs in fewer than 1\% of the tests. Some answers do not provide a numerical value: They may provide a range ({\tt A fair hourly wage in US\$ for a materials worker would be between \$13 and \$15 per hour}), or they may refuse to provide a number altogether.

\noindent In some cases, the verbal response appears to corroborate the numerical answer:
{\linespread{1}
\begin{quote}
\tt 
Using data from the Bureau of Labor Statistics, the median hourly wage for a material moving worker is \$16.57. 
\end{quote}
}
However, browsing data from the Bureau of Labor Statistics accessible from the web, I did not find a figure specifically for ``material moving worker,'' and inputting the query into Google Search yielded \$14.58/hour for Hand Laborers and Material Movers.
A different  test on the identical prompt with identical setting yielded another inconsistent answer:
{\tt 
The average hourly wage for a materials worker is \$13.50.
}
Some of the most interesting responses were observed for the prompt $m = 0$, where the minimum wage is said to be absent (as opposed to nothing being said about the minimum wage). For example, the following two responses were obtained from identical input and settings, reaching seemingly contrary conclusions on the relation between fair wage and minimum wage:
{\linespread{1}
\begin{quote}
\tt There is no fair hourly wage for this worker in the United States because there is no minimum wage. \\ 
There is no minimum wage in this location, so a fair hourly wage for this worker would be \$0.
\end{quote}
}

\subsection{Limitations of the Experiments}
\label{sec:additional}

In our human subject experiments, we attempted to control spurious variables by running the test at the same time of the day (6PM PST) and restricting the poll to adult respondents. However, there are many dimensions that, if changed, may affect the outcome, including  age, level of instruction, geographic location, income and current profession, political inclination, awareness of the minimum wage in their location, and others. I reserve their analysis for future projects. Unlike other aggregate analyses where GPT-3 can be used at least as a coarse proxy, intersectional analysis cannot be conducted by using an LLMs since the user has no visibility into how the LLM is trained and how attributes of individuals who contributed to the training data affect the trained model. An additional dimension to explore is the type of job. I have chosen the most common jobs in the US for these experiments, which happen to be jobs for which minimum wage considerations apply. I have not tested the effect for jobs that are widely understood as being compensated at a level far above the minimum wage such as physicians, airline pilots, nurses, college professors, etc. I leave this extension for future investigation as well. 

\subsection{Sociopolitical Implications}
Given the results established in this paper, how should governments regulate labor markets through minimum wages? One interpretation is that as long as the minimum wage is below the perceived fair wage, the minimum wage results in decreases in the judgements of perception of fairness, meaning that current minimum wages in the United States, which range from \$7.25/hr to around \$15/hr, may do more harm than good. However, one might respond that perceptions of fairness are by no means the only determinant of wages, and that many people are paid wages significantly below what is considered fair according to this study's respondents. Thus, some may argue that the minimum wage should be increased to avoid an anchoring effect in the downward direction. Although this research does not resolve the question of how the government should set the minimum wage, it adds a new dimension by demonstrating how psychological phenomena can factor into the effects of policy.

%\newpage

%\setcounter{page}{1}

%\printbibliography{}

%\bibliographystyle{unsrtnat}
%\bibliography{Anchoring}

\subsection*{Acknowledgments}

I thank Dr. Kirby Nielsen for providing suggestions early in this project, including recommending the use of the platform {\tt Prolific.co}, and Dr. Pietro Perona for showing me Amazon GroundTruth and providing many comments and suggestions for additional experiments. In particular, after reviewing an earlier draft, Dr. Perona suggested conducting additional experiments with unrealistically high values of the anchor, which I chose to be \$50 and \$100. Dr. Perona also pointed me to recent references that applied crowdsourcing and GPT-3 to social science experiments, and he encouraged me to replicate classical anchoring experiments with the AI models. Dr.  Stefano Soatto suggested the use of Mutual Information for measuring the anchoring effect in Sect.~\ref{sec:scores} and helped create the graph in Fig.~\ref{fig:bimodal}  from data I collected from the experiments. Finally, Dr. Colin Camerer provided me with the seed of inspiration for this study by allowing me to attend his research group meetings during a summer internship in his laboratory.

\begin{figure}[h]
\linespread{1}
\begin{center}
\includegraphics[width=.45\textwidth]{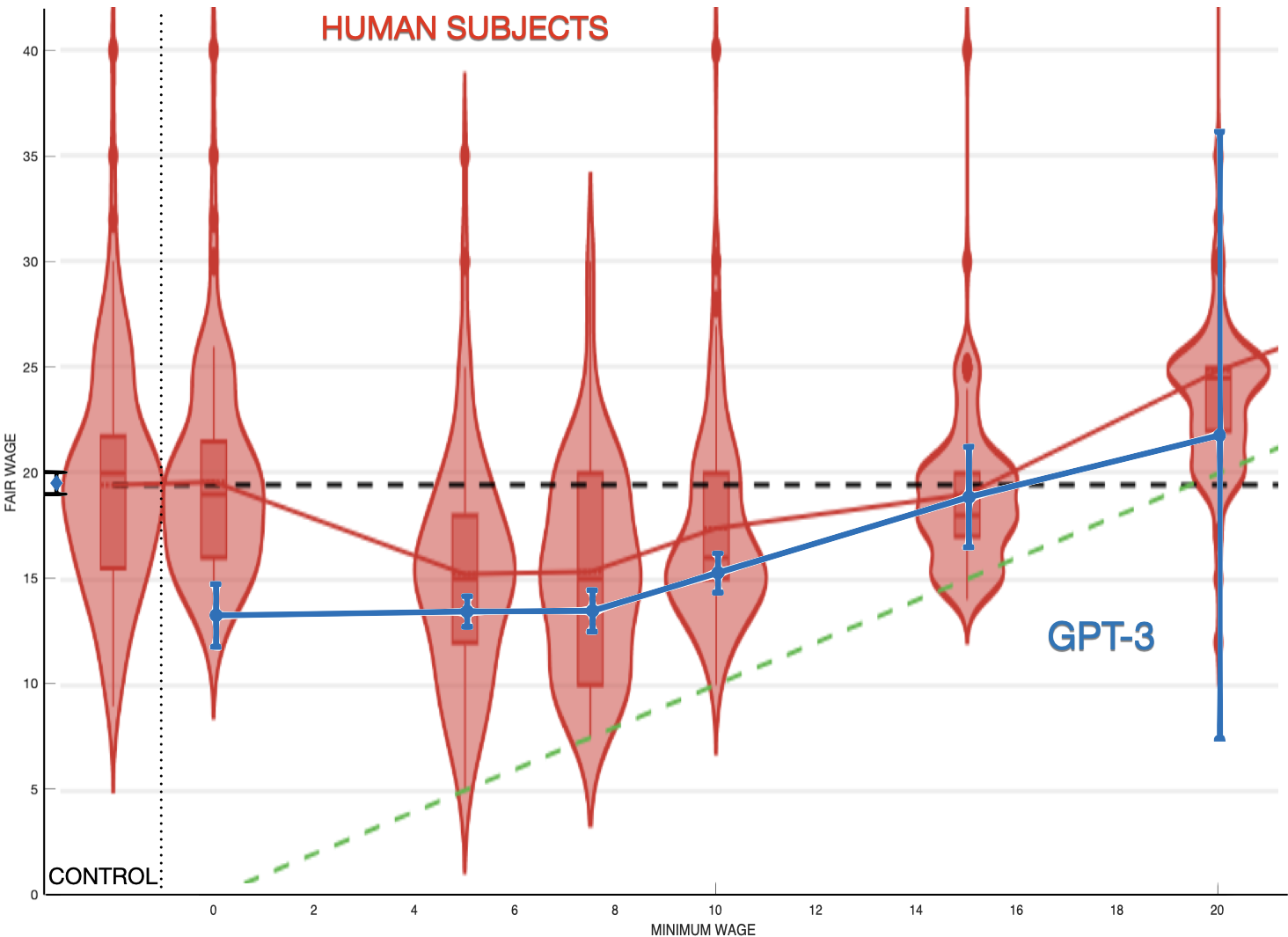}
\hspace{.04\textwidth}
\includegraphics[width=.45\textwidth]{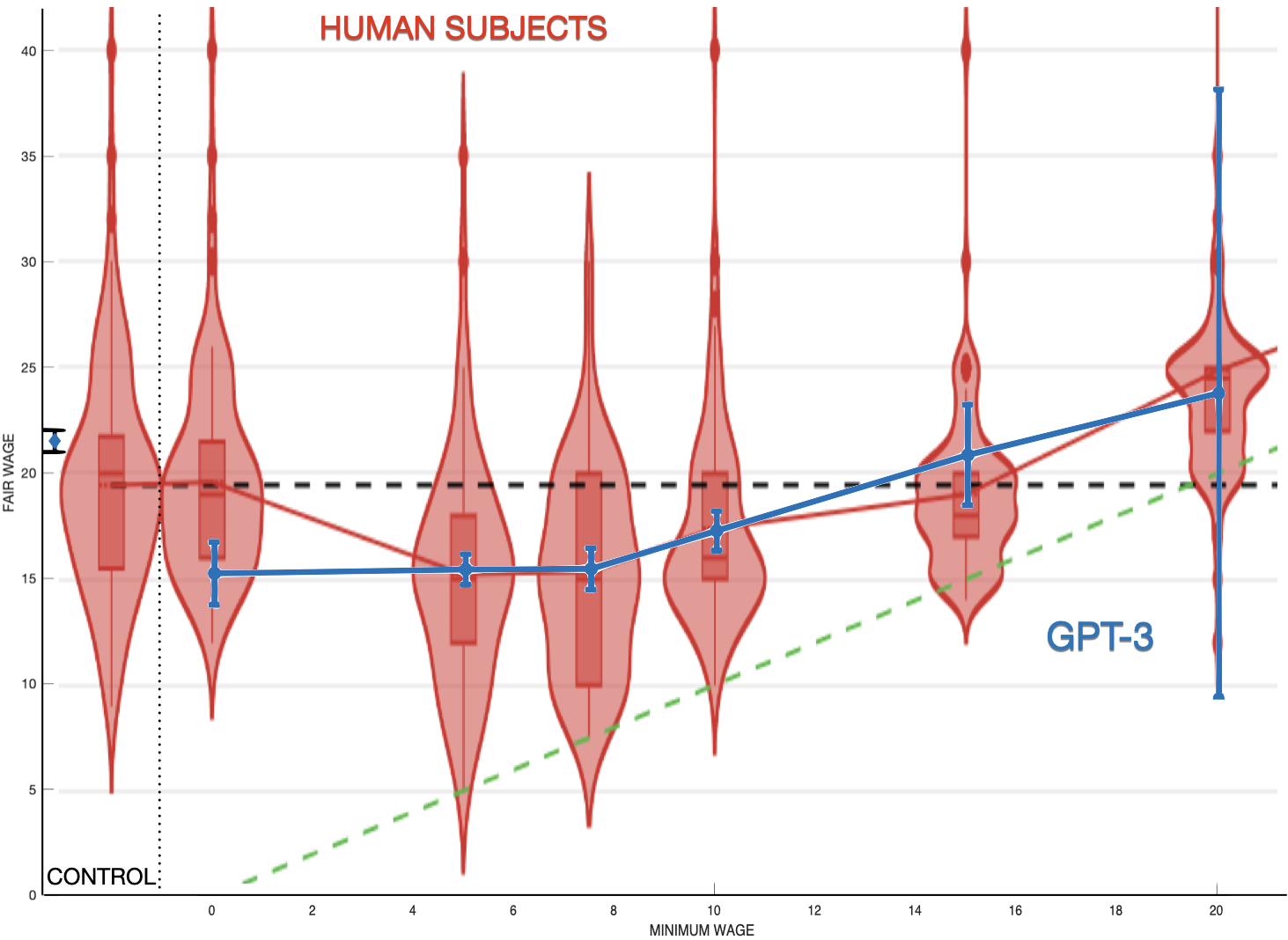}
\end{center}
\vspace{-.3cm}
\caption{{\bf Aligning GPT-3's responses to humans'.} While Fig.~\ref{fig:splash} shows the raw data, here we shift GPT-3's response to align the control to humans' (left panel) and to align the average response to humans' (right panel), which corresponds to a shift upwards by \$5.32.
}
\label{fig:aligned}
\end{figure}

\end{document}